**Adamantane plasma polymers: fluorine-free vacuum-processable triboelectric thin films for all-triboelectric nanogenerator configurations**

*Gloria P. Moreno-Martínez, Fernando Núñez-Gálvez, Hari Krishna Mishra, Triana Czermak, Xabier García-Casas, Vanda Cristina Godinho, Bernd Wicklein, Juan Carlos Sánchez-López, Javier Ferrer, Isabel Montero, Juan Ramón Sánchez-Valencia, Andris Sutka, Francisco Aparicio, Angel Barranco,* Ana Borras**

G. P. Moreno-Martínez, F. Núñez-Gálvez, H. K. Mishra, T. Czermak, X. García-Casas, V. C. Godinho, F. J. Aparicio, J. R. Sánchez-Valencia, A. Barranco, A. Borras
Nanotechnology on Surfaces and Plasma Laboratory, Consejo Superior de Investigaciones Científicas (CSIC), Materials Science Institute of Seville (CSIC-University of Seville), Seville, Spain.
E-mail: anaisabel.borra@icmse.csic.es, angelbar@icmse.csic.es

F. Núñez-Gálvez
Departamento de Física Aplicada I, Universidad de Sevilla, Seville, Spain.

B. Wicklein, I. Montero
Materials Science Institute of Madrid (ICMM), Consejo Superior de Investigaciones Científicas (CSIC), Madrid, Spain.

J.C. Sánchez-López
Tribología y Protección de Superficies, Consejo Superior de Investigaciones Científicas (CSIC), Institute of Materials Science of Seville (CSIC-US), Seville, Spain.

F. J. Ferrer
Centro Nacional de Aceleradores, Universidad de Sevilla, Seville, Spain.

A. Sutka
Institute of Physics and Materials Science, Faculty of Natural Sciences and Technology, Riga Technical University, Riga, Latvia.

Funding: The authors thank the projects PID2022-143120OB-I00, PCI2024-153451, and TED2021-130916B-I00, funded by MCIN/AEI/10.13039/501100011033 and by "ERDF (FEDER)" A way of making Europe, Fondos NextGenerationEU, and Plan de Recuperación, Transformación y Resiliencia and Project 202460E235 funded by CSIC. Project ANGSTROM was selected in the Joint Transnational Call 2023 of M-ERA.NET 3, an EU-funded network of about 49 funding organisations (Horizon 2020 grant agreement No 958174). FNG acknowledges the "VII Plan Propio de Investigación y Transferencia" of the Universidad de Sevilla. The authors further acknowledge the SUsPlast platform (CSIC). GPM thanks the MICIN/AEI/10.13039/501100011033 for the FPI Grant (State Programme for the Promotion of Talent and its Employability in R+D+I, PRE2020-093949). The project leading to this article has received funding from the EU H2020 program under grant agreement 851929 (ERC Starting Grant 3DScavengers).

Triboelectric nanogenerators (TENGs) are major drivers in on-site power generation for smart devices, enable self-powered sensors, and introduce novel catalytic processes. Here, we present the advantages of adamantane plasma layers as bivalently triboelectric surfaces capable of exhibiting both tribopositive and tribonegative character through simple modification of the synthesis conditions without the need for additives or functionalization. Fabrication facing or backfacing the plasma yields thin film polymers with different dielectric constants, Young's moduli, and secondary electron emission. The conformality, stability, and processability of the polymers enable direct implementation across solid-solid, solid-liquid, and hybrid piezo-triboelectric configurations. Additional texturization by buckling is shown to provide voltage and current outputs as high as 90 V $cm^{-2}$ and 0.6 μA for a 2.8 μm (tribonegative) vs. 400 nm (tribopositive) combination. A maximum power density of 2.1 μW $cm^{-2}$ is generated from salty droplets in a switch-electrode drop-TENG configuration employing a 500 nm-thick tribopositive adamantane polymer as the triboelectric surface. These layers have demonstrated outstanding durability, enabling more than $10^5$ cycles in solid-solid nanogenerators and $10^4$ droplet impacts in solid-liquid configurations. The synthetic method is environmentally friendly and industrially scalable, making the adamantane plasma polymer a reliable and competitive solution for thin film triboelectric materials.

## 1. Introduction

The field of triboelectric nanogenerators (TENGs) has grown exponentially since the first report was published by Wang et al. in 2012.[1] In their most extended form, these nanogenerators enable direct conversion of low- and mid-frequency vibrations by exploiting contact electrification and electrostatic induction that occur when two dissimilar interfaces are brought into contact.[2] A significant advantage of TENGs over other energy-harvesting systems is that the conversion mechanism does not depend on defined crystalline phases or orientations (which is of utmost importance), unlike piezoelectric and pyroelectric nanogenerators. Thus, TENGs can be engineered by combining surfaces with different electron affinities, following, for instance, the trends settled in the triboelectric series.[3–5] The ample variety of combinations of polymers (Teflon, polyvinyl chloride), metals, natural materials (cellulose, natural rubber, silk), synthetic surfaces, and the flexibility of device architectures have expanded the scope of nanogenerators in terms of technological merit. The integration of the various operational modes (vertical contact separation, lateral sliding, single-electrode and free-standing electrode modes)[6,7] and interfacial surfaces (solid-solid, solid-liquid, and liquid-liquid interfaces)[8] has fostered the applications of these nanogenerators beyond microscale kinetic energy harvesting for wearable electronics. These applications include large-scale conversion of blue energy (wind, water, rain),[9] self-powered touch sensors,[10] advanced self-powered sensors,[11] for health and environmental monitoring and smart agriculture,[12] air purification,[13,14] electrostatic discharge protection,[15] development of anti-fouling active surfaces,[16] and activation of catalytic processes.[17]

However, three gaps must still be addressed in this appealing scope before TENGs can be fully realized and implemented. First, tribonegative materials are primarily composed of fluoropolymers, such as polytetrafluoroethylene (PTFE), poly(vinylidene fluoride) (PVDF), and fluoroelastomers (FKMs), which require perfluorinated and polyfluorinated alkyl substances (PFAS) for their production.[18] These molecules are commonly referred to as "forever chemicals" owing to their persistence and environmental concerns, and are increasingly subject to regulatory action, particularly in Europe, with further restrictions expected in the coming years. Finding alternative polymers with similar chemical and mechanical stability, and surface and electrical performance is an urgent challenge to be addressed by materials scientists across disciplines, including microelectronics, protective coatings, antifouling and antimicrobial technologies, and smart wetting surfaces.[19] Secondly, the synthesis of polymeric materials, regardless of their chemical composition, should comply with green chemistry and material-reduction principles. This ensures high-yield production of reproducible and finely tailored surfaces at low environmental and energy costs. Finally, the new generation of triboelectric polymers must be adaptable to the multiple configurations of triboelectric nanogenerators (solid-solid or solid-liquid interactions), including direct compatibility with micro- and nanostructuration and interfacial adaptation to hybrid systems, such as combinations with piezoelectric or ferroelectric fillers or ferroelectric porous architectures.

In this article (see **Scheme 1**), we propose to address these challenges by exploiting a novel family of plasma polymer thin-film materials enabled by an evolved version of plasma polymerization, termed remote plasma-assisted vacuum deposition (RPAVD).[20] The basis of the process is the sublimation of small, vacuum-processable functional molecules in the downstream region of a remote cold plasma (Scheme 1, Top). The regulated plasma interaction induces the formation of a non-soluble, cross-linked matrix, while minimizing precursor fragmentation, preserving functional groups, and even enabling the controlled retention of a fraction of intact molecules within the film. RPAVD is a solvent-free and industrially scalable technique that profits from the inherently low-temperature, low-power nature of plasma-activated processes. The technique is straightforwardly compatible with other vacuum and plasma deposition and processing methods, as well as with soft-hard templating, etching, laser engraving, and photolithography (Scheme 1, Bottom).

The deposition is carried out downstream (out of the glow discharge), making it highly compatible with the conformal formation of layers over delicate substrates, such as organic surfaces and cellulose, and three-dimensional micrometric structures, including organic nanowires, 2D materials and photonic and optoelectronic structures.[21,22] This fabrication applies to a variety of small molecules that are processable under vacuum, such as rhodamines,[23] flavonols,[24] DCM,[25] perylenes,[26,27] phthalocyanines,[28] and porphyrins.[29,30]

Among these molecules, adamantane is the simplest diamondoid molecule, a solid hydrocarbon composed of $sp^3$ carbon arranged in a cage-like structure, with formula $C_{10}H_{16}$, obtained from natural sources.[31] RPAVD Adamantane plasma layers have already demonstrated enhanced properties as conformal anti-freezing coating,[32] for device encapsulation and passivating interfaces of halide perovskite solar cells [25], [26], [27], and for strain engineering of 2D materials, such as $MoS_2$ [28].

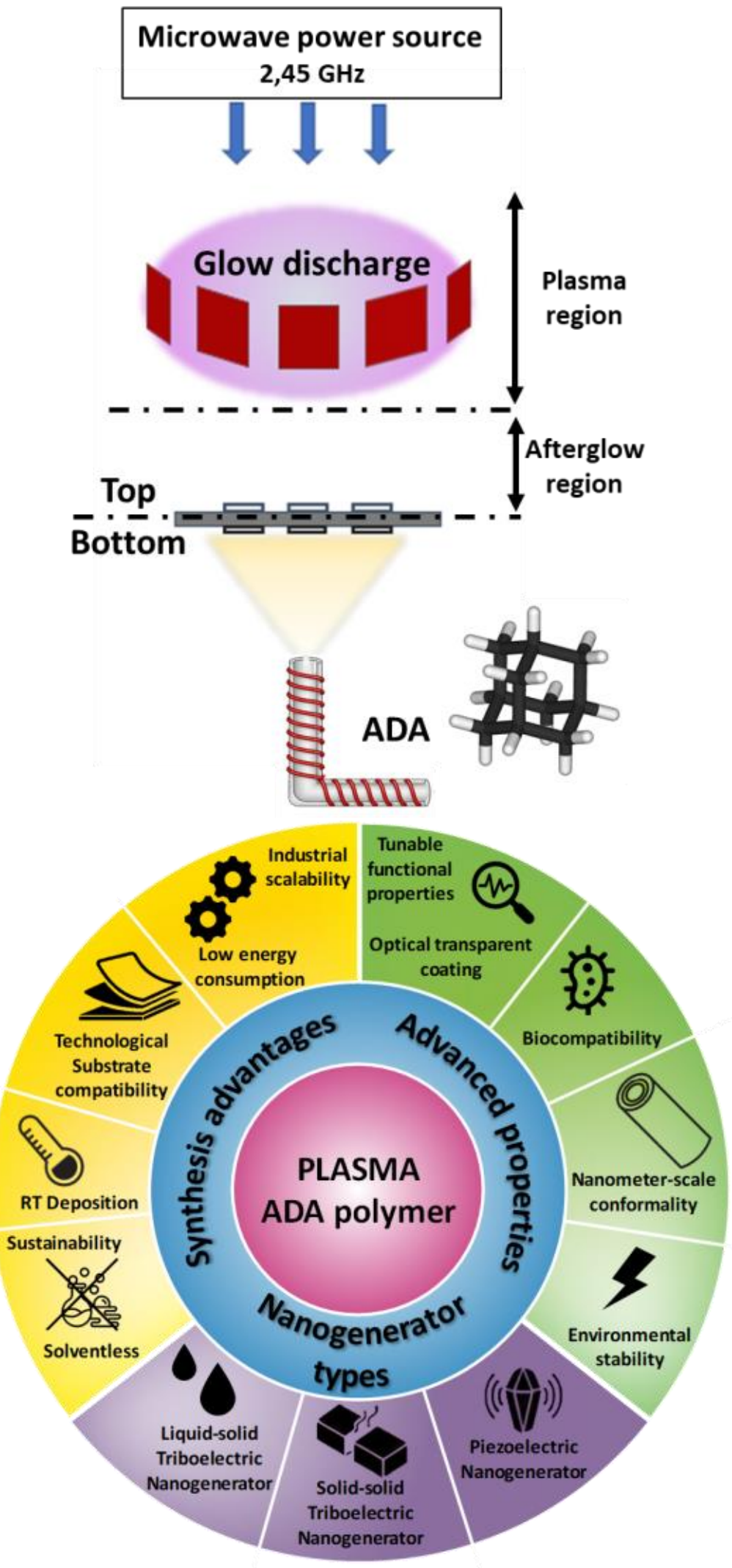

**Scheme 1.** Top) Schematic representation of the remote plasma-assisted deposition of polymer-like layers, particularly for the fabrication of adamantane layers facing up (Top-ADA) and down (Bottom-ADA) to the plasma discharge. Bottom) Applications and scope of the implementation of adamantane plasma polymers in TENGs.

Herein, we take a step further by demonstrating their outstanding performance as triboelectric layers across all nanogenerator configurations and by tuning their triboelectric character through a facile modification of the plasma precursor interaction. We first present the physicochemical properties of the layers in relation to the fabrication conditions, particularly the substrate orientation,[33] facing up (higher crosslinked material) or facing down toward the plasma discharge. Secondly, we analyze the triboelectric character and performance of these layers in vertical-contact-mode solid-solid TENGs to establish their position in the triboelectric series and to demonstrate tunability from tribopositive to tribonegative polymers, matching the power outputs of bulk fluorinated counterparts in a buckling configuration while exhibiting high durability. In the final part, we implement adamantane in microstructured TENGs as a proof-of-concept triboelectric counterpart of a mode-I hybrid piezo-triboelectric nanogenerator and a solid-liquid nanogenerator, specifically as a friction layer embedded between an extended bottom electrode and a thin top electrode (switch electrode layout [34]) for water droplets energy harvesting.

## 2. Results and discussion

### 2.1. Adamantane thin film deposition and properties.

In the previous examples of RPAVD,[35] the most commonly used deposition conditions involve placing the substrates facing down in the under-glow region of a microwave electron cyclotron resonance (MW-ECR) plasma discharge (Scheme 1, Bottom-ADA position) to reduce the plasma interaction with the growing film. Here, we present, for the first time, the properties of films prepared facing towards the plasma discharge, i.e., Top-ADA layers. The dynamics of fragmentation and cross-linking of the material growing as Top-ADA are expected to be closer to those of a more conventional plasma polymerisation process, although in remote conditions and at low MW plasma power, as the plasma interaction with precursor and growing film increases in this position. Notably, the growth rate under Top-ADA deposition conditions is approximately five to six times higher than under Bottom-ADA conditions. Henceforth, we will either compare the properties and performance of the Bottom and Top layers prepared during the same synthetic experiment (i.e., when the top layer is thicker than the bottom layer) or purposely prepare the top and bottom layers with the same thickness in separate experiments. To facilitate reading of the article, we include in **Table 1** a summary of the sample labelling, indicating the position in the holder (Scheme 1) and distinguishing smooth films from buckled layers, the latter obtained for thick films deposited facing the plasma on low surface adhesion substrates (e.g., ITO and gold).

**Table 1.** Labelling of the triboelectric layers, attending to the deposition geometry and morphology.

| Adamantane surface | Geometry of the deposition | Thickness (nm) | Morphology |
|---|---|---|---|
| Top-ADA | Top position, facing plasma discharge | 450-600<br>Typically 500 nm | Smooth (flat, low roughness configuration) |

| | Si, glass, FTO-glass, ITO-PET | | |
|---|---|---|---|
| Bottom-ADA | Bottom position, downfacing plasma discharge<br>Si, glass, FTO-glass, ITO-PET | 450-600 nm<br>Typically 500 nm | Smooth (flat, low roughness configuration) |
| Top-ADA buckled | Top position<br>ITO-PET, Au-Si | > 1000 nm<br>Typically 2800 nm | Buckled (low roughness at the nanoscale, buckling on low-surface adhesion substrates) |

The overall composition has been studied using combined proton- Elastic Backscattering Spectroscopy (p-EBS) and Elastic Recoil Detection Analysis (ERDA) measurements of films deposited on Si(100). The results are summarised in **Table S1 in the Supporting Information Section**. The top and bottom films show C and H atomic percentages very close to those of the adamantane molecular formula ($C_{10}H_{16}$), with H contents of ca. 60%. In contrast, both film types exhibit minor oxygen content of 5.4% and 2.1% for the Top-ADA and Bottom-ADA samples, respectively. Oxygen incorporation is a common characteristic of plasma polymerisation processes. The higher oxygen content in the top sample can be attributed to stronger interaction with plasma species during synthesis, which generates a higher density of radicals that subsequently react with oxygen upon exposure to air, as well as to the incorporation of reactive oxygen-containing species inherently present in the plasma.[26,36]

The Raman spectra of the films (**Figure 1**a) exhibit relatively low-intensity bands, with a noticeable contribution from atmospheric $N_2$ (2325 $cm^{-1}$, marked with an asterisk). The more intense Raman bands at ~2925 $cm^{-1}$ are assigned to aliphatic C-H stretching modes characteristic of saturated $sp^3$ carbon.[37] In contrast, the less intense and broad feature ~1608 $cm^{-1}$ indicates a contribution from $sp^2$-bonded carbon (C=C stretching / G-like mode).[38,39] The intensity ratios (I(2925 $cm^{-1}$)/I(1608 $cm^{-1}$)) are 2.92 and 1.86 for the Bottom-ADA and Top-ADA, respectively. These ratios indicate less intense precursor fragmentation and cross-linking in the Bottom-ADA samples, consistent with reduced plasma interaction during growth, as expected from the deposition geometry (see Scheme 1, Top/Bottom configurations).

SEM and AFM characterisations (**Figure S1**) show that the layers exhibit a compact, homogeneous cross-section and a flat surface. The RMS surface roughness for both configurations at the equivalent thickness is extremely low; however, Top-ADA RMS (1.5 nm for 5 x 5 $\mu m^2$) is more than twice that of Bottom-ADA (0.6 nm for 5 x 5 $\mu m^2$). Extremely low surface roughness in films ranging from tens to hundreds of nm has been previously reported for conformal films synthesized by RPAVD from different functional precursors (i.e., under Bottom-ADA experimental conditions).[23,32,35,36]

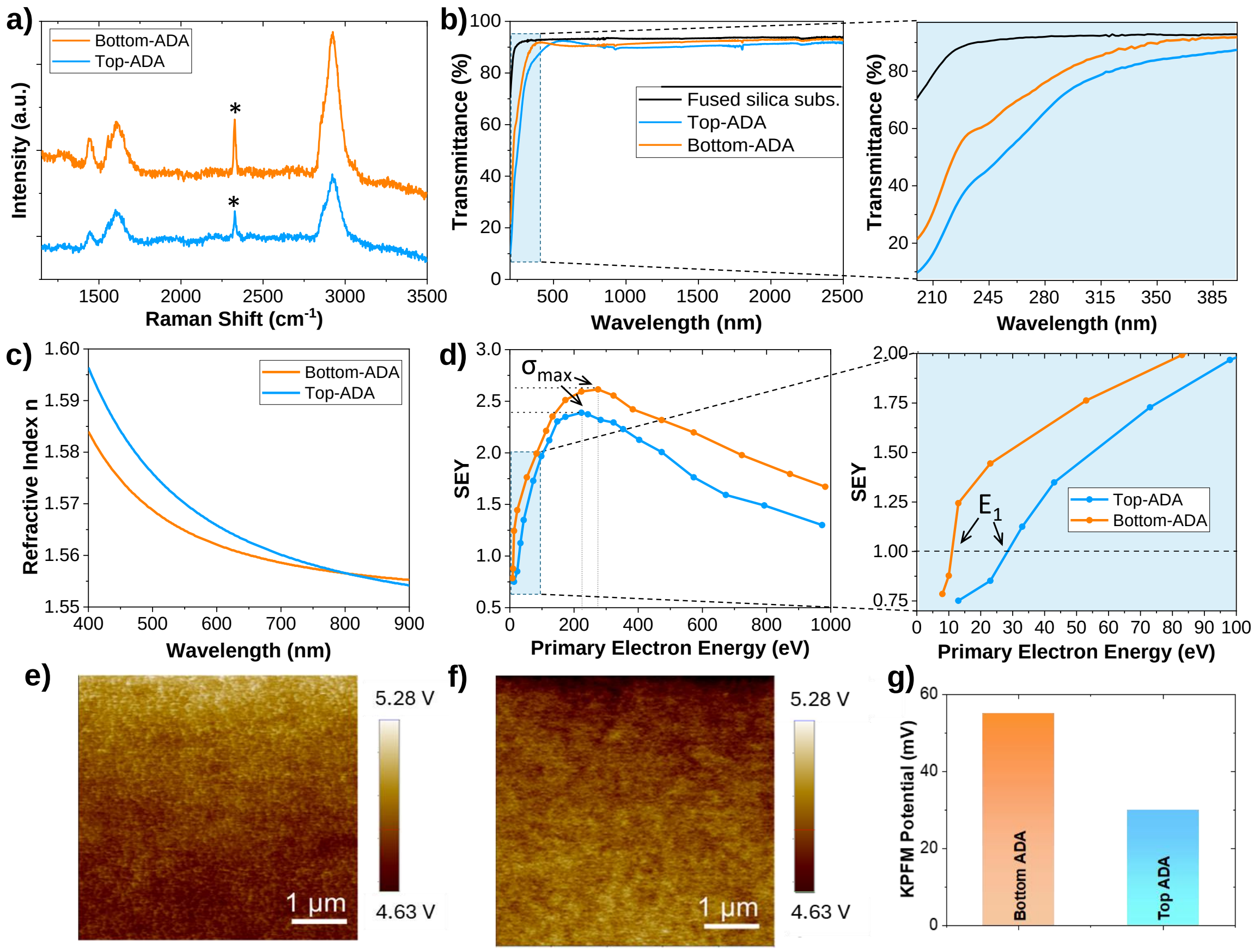

**Figure 1. Characterisation of Top (orange) and Bottom (blue) ADA films.** a) Raman spectra (the asterisk indicates atmospheric $N_2$). b) UV-vis-NIR transmittance spectra including the fused silica substrate in black. The curve on the right side corresponds to a magnified view of the area marked in blue. c) Refractive index curves determined by VASE. d) Secondary Electron Yield (SEY) spectra of the films. The curve on the right side of d) corresponds to a magnified view of the area marked in blue. The KPFM surface potential maps for e) Bottom-ADA and f) Top-ADA thin films. g) KPFM contact potential difference of both thin films.

ADA Top and Bottom films are transparent across the UV-vis-NIR regions, as shown in Figure 1b. Bottom-ADA films are slightly more transparent than Top-ADA films in the UV region (see highlighted view on the right side of the panel). The refractive indices of both types of films, determined by variable angle spectroscopic ellipsometry (VASE), are similar, with the Top-ADA films slightly higher ($n_{550nm}$=1.570) than the Bottom-ADA films ($n_{550nm}$=1.565).

These differences in the optical properties of the films are minor but highly reliable and reproducible, as observed across films deposited under identical conditions. Besides, the relative permittivity of the films is 3.5 and 2.75 for the Top and Bottom-ADA configurations, respectively. This difference is more pronounced than for the optical values.

Similarly, the mechanical properties are also affected by the experimental growth conditions. It has been shown that mechanical properties are crucial to the performance of triboelectric materials and their charging tendencies. Thus, for bulk and fiber counterparts, softer polymers tend to charge negatively while harder polymers charge positively.[40] Moreover, softer materials present higher charging capability, attributed to a higher specific contacting area.[41] **Table S2** in the Supporting Information presents the Young elastic modulus (E) and hardness (H) values obtained from nanoindentation curves using the Oliver-Pharr method for equivalent thin-film Top-ADA and Bottom-ADA prepared to a thickness of approximately 500 nm. The resulting values are E/H = 10.4 ± 0.5 GPa / 0.75 ± 0.06 GPa for Top-ADA, and 7.5±0.3 GPa / 0.55 ±0.06 GPa for Bottom-ADA. Typical tribonegative flourinated polymers such as PTFE and PFA exhibit indentation moduli below 1 GPa.[42,43] Similarly, plasma polymer coatings and conformal coating materials such as $CF_x$ and parylene typically exhibit elastic moduli in the range of 0.1–5 GPa [44,45] and 3-4 GPa,[46] respectively, which remain significantly lower than those measured for Top-ADA and Bottom-ADA films. Tribopositive polymers such as semicrystalline nylon and amorphous polycarbonate also display substantially lower values, typically in the range of 2-3 GPa.[47] More generally, conventional structural thermoplastic polymers such as PMMA, PC, PET or PEEK typically exhibit elastic moduli between 2 and 5 GPa, depending on morphology and processing conditions.[42,48] The measured values are also higher than those commonly reported for epoxy crosslinked polymer networks and comparable to or greater than those of unreinforced polybenzoxazine (PBZ) systems, approaching the stiffness of rigid polyimide networks.[49,50] These results indicate the formation of a densely crosslinked glassy polymer structure with outstanding mechanical features among high-performance polymeric materials.

Thus, the RPAVD ADA coatings represent a compromise among a relatively high Young's modulus, a fully organic composition, complete optical transparency extending into the UV range. Moreover, unlike plasma-based amorphous carbon coating deposition, in which energetic ion bombardment governs film densification and $sp^3$-rich network formation, RPAVD synthesis enables conformal deposition on delicate substrates and on molecular nanostructures.[22,35,39]

Secondary electron yield (SEY) measurements, defined as the ratio of the emitted secondary-electron charge to the incident primary-electron charge as a function of incident electron energy, are widely used to characterise the charge-emission behaviour of materials under electron irradiation.[51] The analysis of the SEY curve over a broad energy range allows for the prediction and modelling of electron-induced phenomena such as surface charging, electron multiplication, and discharge processes in vacuum environments. SEY measurements are sensitive to the effective electronic and chemical properties of the near-surface region that govern secondary-electron generation, transport, and escape. Hence, SEY measurements play a critical role in the design and assessment of materials for applications involving electron transport and irradiation, including dielectric coatings, polymeric films, and plasma-processed surfaces. Figure 1d shows the SEY curves of the Top-ADA and Bottom-ADA films, exhibiting the characteristic SEY behavior with a broad maximum, together with clear differences in yield magnitude and peak position. Increasing plasma polymerisation results in a systematic shift of the SEY parameters, with both the maximum yield ($\sigma_{max}$) and the corresponding primary electron energy ($E_{max}$) decreasing from 2.62 at 265 eV to 2.38 at 220 eV. Across the entire investigated energy range, the Bottom-ADA film displays a higher SEY than the Top-ADA film. This behaviour is attributed to reduced effective electron scattering and an enhanced secondary-electron escape probability, associated with the lower cross-linking density of the plasma-polymerised structure. The $E_{max}$ and $\sigma_{max}$ values are comparable to those reported for technologically relevant dielectric polymers such as Kapton (250-323 eV / $\sigma_{max} \approx 2.7$-2.9), Teflon (300-473 eV/$\sigma_{max} \approx 2.0$-2.1), PMMA (≈273 eV/$\sigma_{max} \approx 2.7$-2.9), Nylon (≈273 eV/$\sigma_{max} \approx$ 2.7-2.8), and polyethylene (≈273 eV/$\sigma_{max} \approx 3.1$).[51,52]

A more pronounced distinction between the two films is observed in the low primary-electron-energy region, shown in the right panel of Figure 1d. The first crossover energy $E_1$, defined as the primary electron energy at which the SEY equals unity (i.e., the sample emits the same charge it receives), is 11.1 eV for the Bottom-ADA film and 28.4 eV for the Top-ADA film. Although the $E_1$ value of the Top-ADA sample is higher than that of the Bottom-ADA, both remain far from those reported for polymers such as PMMA (43 eV), polyethylene (34 eV), Nylon (35 eV), and Teflon (33 eV), and significantly lower than that of Kapton (123 eV).[52] The remarkably low $E_1$ value observed for the Bottom-ADA film indicates a more efficient secondary-electron emission at very low excitation energies, revealing a clear electronic distinction between the two plasma-polymerised layers. In these less cross-linked films, secondary electron emission becomes effective at lower primary energies, reflecting a reduced effective surface barrier for electron escape from the near-surface region. $E_1 \sim \phi - \chi$, where $\phi$ is the surface work function and $\chi$ the electron affinity. The low $E_1$ of Bottom-ADA suggests a very low, or possibly negative, electron affinity, consistent with its less cross-linked near-surface structure, which facilitates secondary electron emission. In contrast, the Top-ADA film exhibits a higher $E_1$ = 28.4 eV, in line with conventional polymeric dielectrics such as PMMA or Nylon. These results highlight a distinctive surface electronic environment in Bottom-ADA that promotes efficient secondary electron emission and potentially reduces charge accumulation under electron irradiation. This behaviour is consistent with the low electron affinity reported for adamantane-based diamondoids.[53] The implications of this characteristic for interfacial charge-transfer phenomena will be further discussed in Section 2.6.

Additionally, to corroborate the aforementioned analysis of surface charge tunability, Kelvin Probe Force Microscopy (KPFM) was employed to map the localized surface potential of the RPAVD-deposited Bottom-ADA and Top-ADA thin films on ITO-PET substrates. By resolving the contact potential difference ($V_{CPD}$) at the nanoscale, we elucidated the electrical potential characteristics of both Bottom-ADA and Top-ADA surfaces, which are essential to interfacial charge-transfer efficiency in TENGs. Figures 1e and 1f show the KPFM surface potential map for Bottom-ADA and Top-ADA thin films, revealing a highly uniform potential distribution across the scanned area (5×5 $\mu m^2$). The estimated $V_{CPD}$ indicates that the Bottom-ADA exhibits a significantly higher KPFM potential (~55 mV) than the Top-ADA (~32 mV), as shown in the histogram in Figure 1g. This distinct gradient in surface potential is a critical prerequisite of the electronic state density and the effective surface charge density for triboelectrification. The higher potential observed in the Bottom-ADA film suggests greater electron-donating capability and the most favourable alignment of surface states to trap charges during mechanical activation.[54,55]

**2.2 Solid-solid triboelectric nanogenerators, triboelectric series.**

In the first step to assess the triboelectric properties of adamantane layers, we assembled various TENG devices using intermittent vertical-contact layouts.[4,6,7] A detailed description of these devices is provided in the Experimental Section. **Figure 2**a-f compares the triboelectric output for devices assembled with Top- and Bottom layers of ADA using cellulose, a well-known tribopositive surface.[56] Quantitative results can be extracted from these comparisons. Thus, the mean power produced by the smooth Top-ADA is four times that of the Bottom-ADA devices; indeed, the smooth Top-ADA peak-to-peak voltage is twice that of the Bottom-ADA (Figure 2b and e). These findings indicate a relatively higher tribonegative character for the Top-ADA and a comparatively tribopositive character for the Bottom-ADA.

However, the most compelling evidence comes from the short-circuit current measurements (Figures 2a and 2d). The Top-ADA generates a current density of 30 nA·cm$^{-2}$, whereas the Bottom-ADA yields a signal barely distinguishable from background noise. Importantly, both layers have comparable thicknesses (450 nm for Top-ADA and 500 nm for Bottom-ADA) and were subjected to identical handling protocols during triboelectric characterisation. The measured results correlate well with the SEY measurements, in which Bottom-ADA shows a lower electron affinity (more positive triboelectric charging tendencies than Top-ADA) and consequently exhibits lower negative charging when in contact with tribopositive cellulose.

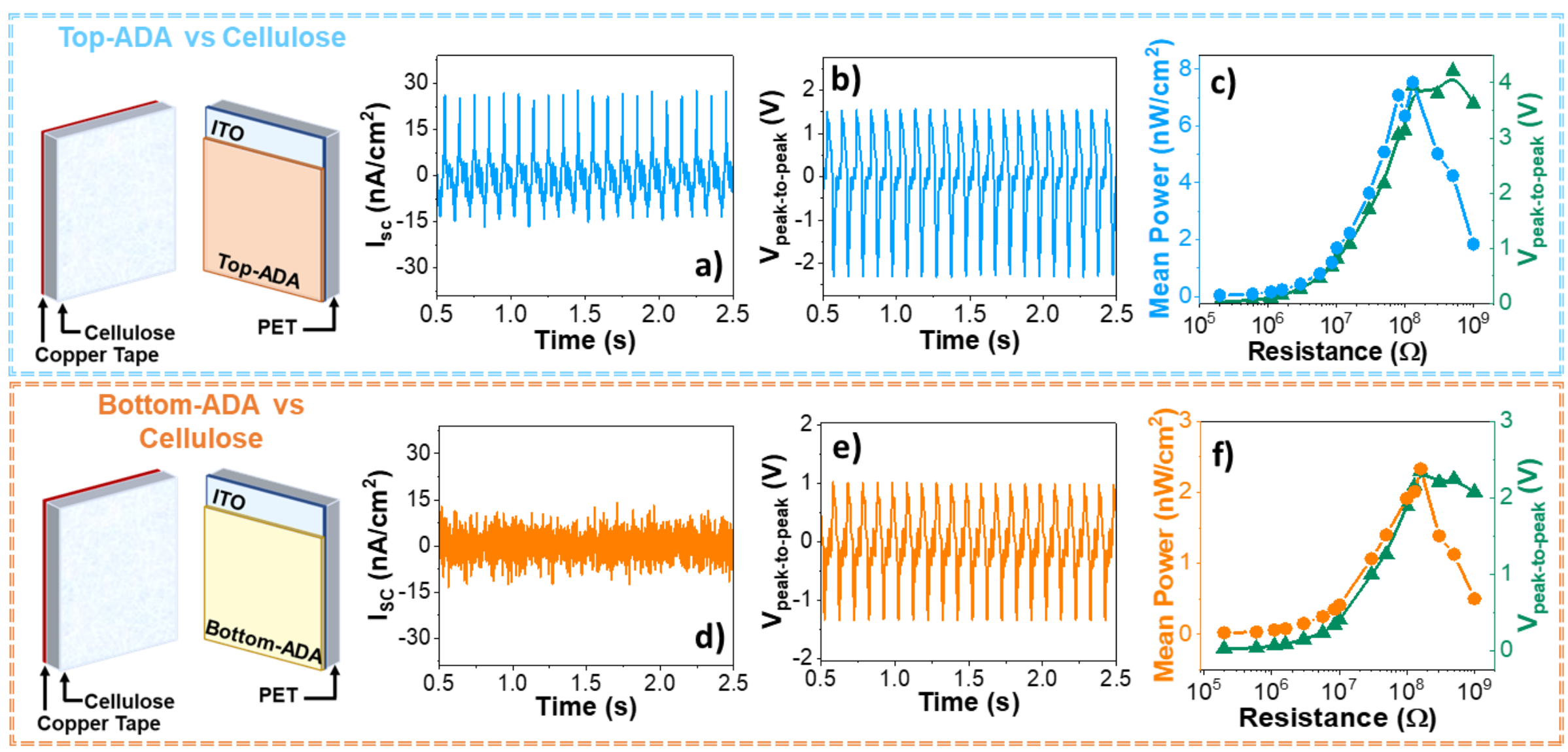

**Figure 2. TENG outputs of RPAVD ADA layers using cellulose counter-electrodes.** Devices assembled with Top-ADA and Bottom-ADA against cellulose as tribopositive surface, as shown in the schematics at the left of the figure. a), d) Short circuit current vs time and b), e) Voltage vs time curves for an input frequency of 10 Hz and a load resistance of 150 MOhm. c), f) Mean power and peak-to-peak voltage output as a function of the load resistance.

To further confirm the triboelectric nature of the Top and Bottom-ADA layers, we systematically tested both triboelectric surfaces against different counter electrodes and triboelectric materials, including PFA, Kapton, ITO-PET, aluminium, and cellulose. **Figure 3** shows the maximum average power for each triboelectric pair (see also supporting information, **Figures S2** and **S3**, for the complete matching impedance data and polarity evolution). Although both Bottom- and Top-Smooth ADA demonstrate tribopositive behaviour, testing against PFA (a representative tribonegative polymer) clearly reveals that Bottom-ADA has a stronger tribopositive character than its top counterpart. These findings are also consistent with the KPFM surface potential measurements (Figure 1g).

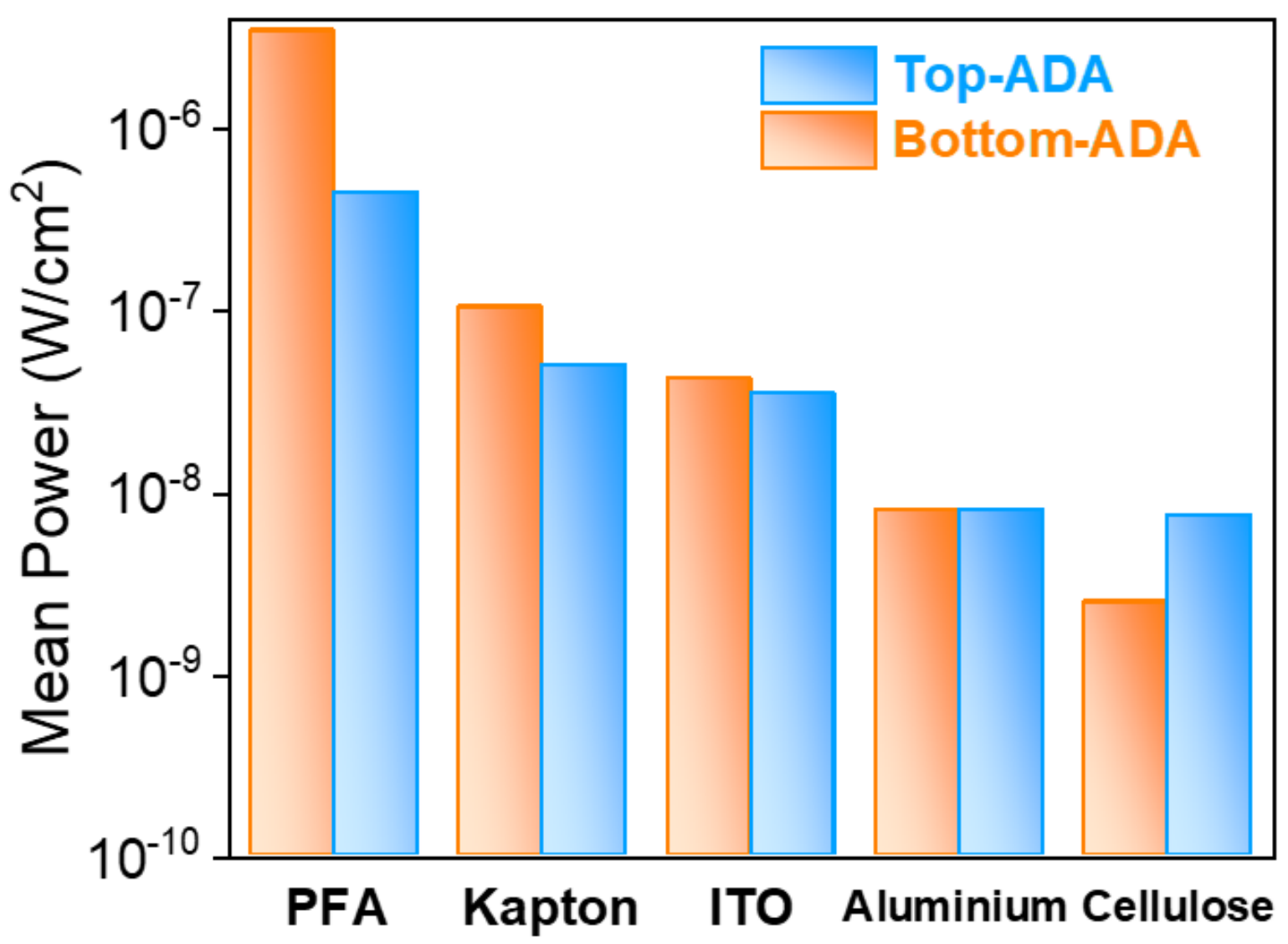

**Figure 3.** Maximum mean power output for Top-ADA (blue) and Bottom-ADA (orange) in intermittent contact mode against a series of counter tribomaterials ranging from tribonegative (PFA) to tribopositive (cellulose) evaluated at 2 N and 10 Hz, under the corresponding impedance matching conditions.

### 2.3 Effect of adamantane thickness on triboelectric performance: Buckled Top-ADA.

The thickness of the adamantane layers is not expected to affect the physicochemical properties discussed in Section 2.1. However, particularly for the Top-ADA configuration, thickness plays an important role in triboelectric performance, as shown herein. For thick Top-ADA films (thicknesses higher than 1 µm, hereafter designated as buckled Top-ADA), factors such as the relatively high deposition rate, the resulting internal stress within the film, and the mismatch in mechanical properties at the film-substrate interface (as in the case of ITO-PET)[57] lead to the appearance of buckling, which strongly determines the surface morphology of the layers. As shown in the optical and SEM images in **Figure 4**, the adamantane layer is partially detached, with localized adhesions to the substrate, resulting in a wavy surface with micrometric gaps between the inner interface of the adamantane film and the exposed substrate surface. Consistent with the results discussed in Figure 3, buckled Top-ADA (2.8 µm thick) is tribopositive when paired with PFA, as presented in **Figure 5**a (see **Figure S5** for the complete triboelectric series). However, in this case, the increase in the frequency of intermittent contact negatively affects the mean power generated. Surprisingly, when buckled Top-ADA is tested against cellulose, the microstructured surface significantly enhances the triboelectric output compared to its smooth counterpart (compare Figure 5b and Figure 2a-c). The mean power increases by a factor of 100, while both peak-to-peak voltage and short-circuit current improve by an order of magnitude. These performance gains would indicate that buckled Top-ADA is a highly effective tribonegative material.

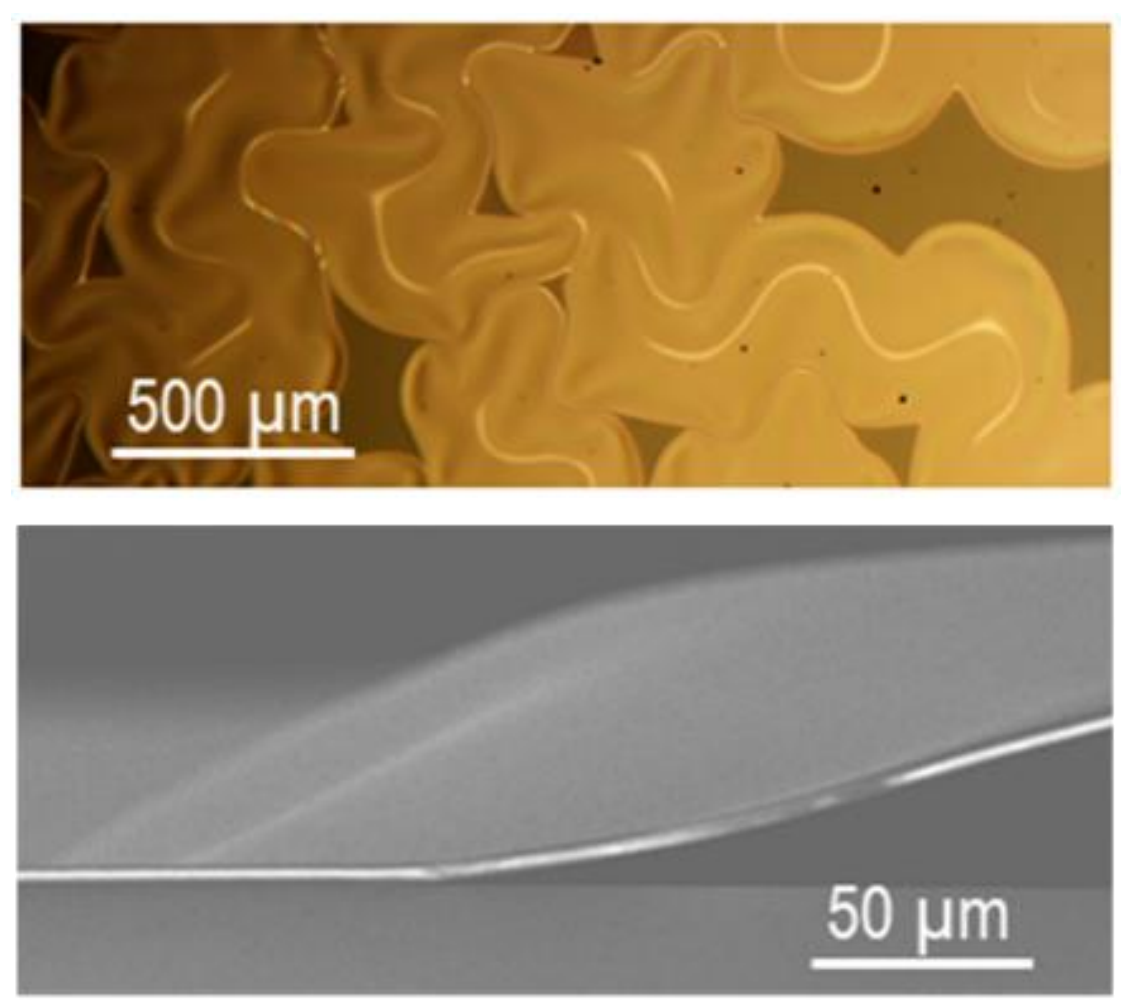

**Figure 4.** Optical image (top) and cross-section SEM micrograph (bottom) of the buckling-induced morphology for Buckled Top-ADA.

To shed light on this apparent contradiction, Figure 5 presents the comparative performance of different triboelectric combinations, including a reference device assembled with ITO and PFA (panel c), Bottom-ADA vs PFA (panel d), and buckled Top-ADA vs Bottom-ADA (panel e) for three pulse frequencies and a fixed force of 2 N as measured by a force sensor (see Experimental Section). The responses recorded for the reference device (panel c) follow the expected trends, with power output increasing with frequency and matching impedances similar to those reported for vertical intermittent contact layouts based on this and similar combinations.[58,59] Bottom-ADA is the most tribopositive ADA plasma polymer when paired with PFA (Figure 5d), reaching values exceeding 2.5 µW cm$^{-2}$. Compared with the reference system, ITO-PFA (Figure 5c), which has a mean power of ~0.75 µW cm$^{-2}$, the Bottom-ADA-PFA power output is more than 3.3 times higher.

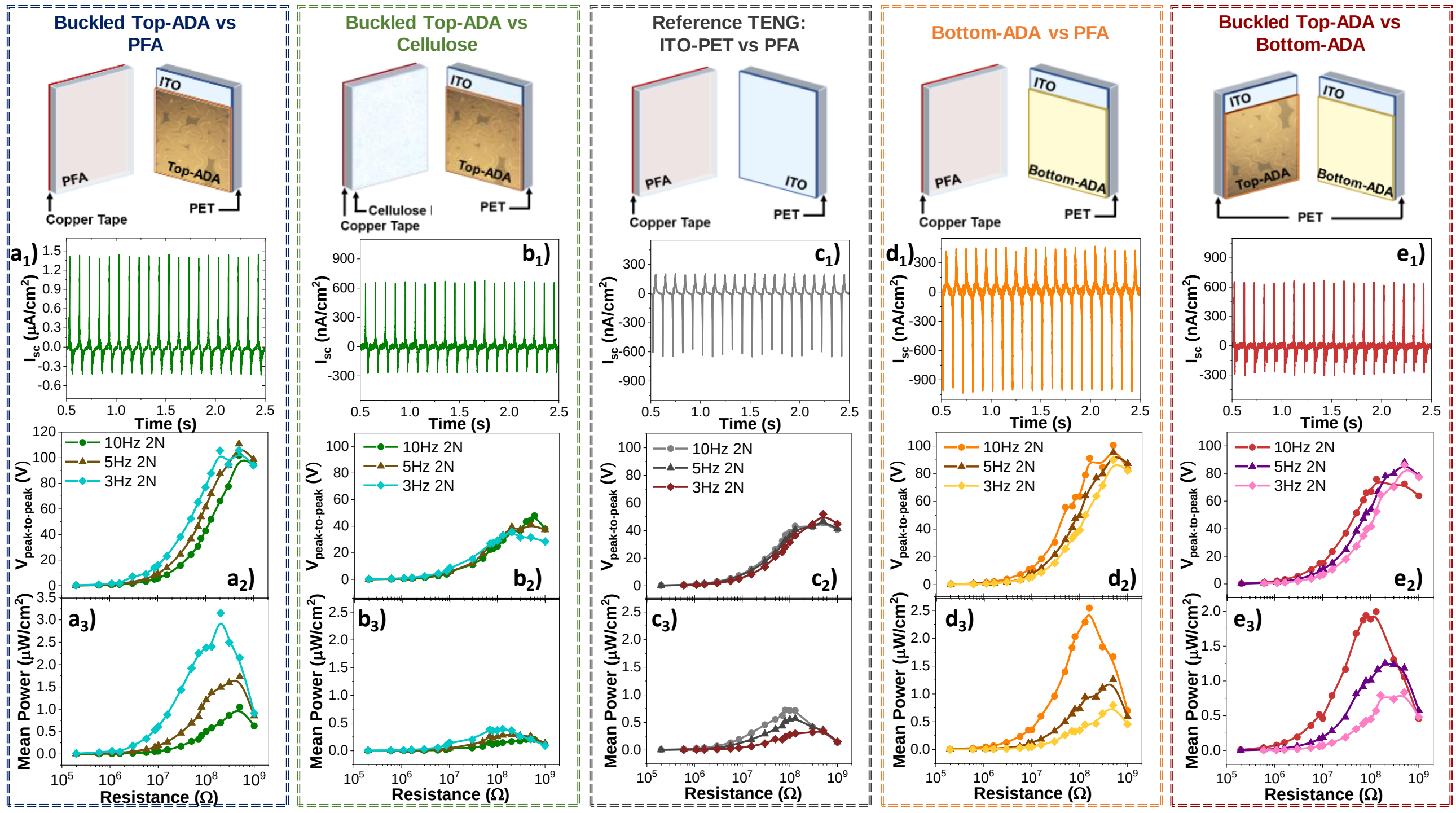

**Figure 5. Buckled Top-ADA triboelectric nanogenerators**. Illustration and a detailed comparison between reference ITO-PFA, Bottom-ADA vs PFA, and buckled Top-ADA (2.8 µm) vs PFA, cellulose, and Bottom-ADA devices, as shown in the schematics at the top of the figure. $a_1$), $b_1$, $c_1$), $d_1$), $e_1$) Short circuit current vs time for an input frequency of 10 Hz; $a_2$), $b_2$), $c_2$), $d_2$), $e_2$) Voltage current vs time curves and $a_3$), $b_3$), $c_3$), $d_3$), $e_3$) Mean power and Peak-to-peak voltage output for varying frequencies as a function of the resistance load.

The performance of the reference pairing was then compared to that of Bottom-ADA coupled with the buckled Top-ADA (Figure 5e). Remarkably, both configurations yielded nearly identical outputs for voltage, current, and mean power. To explain the significant performance improvement in the Bottom-ADA-buckled Top-ADA system, we hypothesize that buckling serves a dual purpose. First, the distribution of valleys and hills increases the surface area, which enhances the contact area for charge transfer between the two surfaces. Previous reports have attributed an enhancement in triboelectric voltage to increasing the surface area and, therefore, the contact area between the triboelectric counterparts. With this aim, references 60 and 61 employed laser patterning to modify the arrangement of microneedles and PDMS. Second, buckling also creates additional gaps between the triboelectric layer and the bottom electrode interface (see Figure 4 bottom). These gaps produce a second TENG structure that is activated simultaneously with the upper surface, thereby enhancing the overall output power. A similar effect was described by M. Zhang et al., in a buckled triboelectric nanogenerator (B-TENG) capable of detecting ultra-slow mechanical stimuli, such as lithium-ion battery swelling.[62] That device exploits a dynamic buckling mechanism, in which both electrodes start flat and buckle during each cycle. This transient deformation significantly enhances the triboelectric response, resulting in a signal up to 200 times greater than that of a flat device. Moreover, the wavy formation can give rise to flexoelectricity,[63] driven by the uneven distribution of deformations between the buckled Top-ADA layer in contact with the underlying electrode and the smooth surface of the Bottom-ADA.

To further elucidate the contributions of these mechanisms, we have produced two additional devices. In the first device, we tested the smooth samples, i.e., the Top-ADA against the smooth Bottom-ADA, achieving a mean power density much lower than that of the buckled counterpart (<3 nW $cm^{-2}$, Figure S5). This result indicates that the enhanced surface morphology or the double interface is a more important factor than the difference in triboelectric character between the top and bottom adamantane layers.

In our second comparative device, we have generated valleys and hills while preventing buckling. **Figure 6**a–c presents characteristic optical and SEM images of the deposition of a ~1µm-thick Top-ADA film through a stainless-steel micromesh wire mask. In this configuration, we can separate the contribution of the increased surface area from that of the secondary triboelectric nanogenerator formed by buckling. The voltage output and mean power for the devices incorporating the patterns in Figure 6f double the output reference (Figure 2b-c). The improvement is also evident in the short-circuit current density (Figure 6e), which reaches 83 nA $cm^{-2}$, i.e., a threefold increase compared to the reference (Figure 2a). However, these values are more than an order of magnitude below those obtained with the buckled Top-ADA configuration (Figure 5d), indicating that although the surface area is increased, the double triboelectric interface induced by buckling dominates the energy-conversion performance in the most efficient combination. Based on this distinction, we hypothesise that the enhanced signal in our device arises not from dynamic buckling velocity but from the static morphological features of the buckled film. Specifically, the presence of micro-gaps between the buckled Top-ADA and the ITO substrate, combined with the surface roughness introduced by the buckling, likely increases local stress and the contact area with the Bottom-ADA.

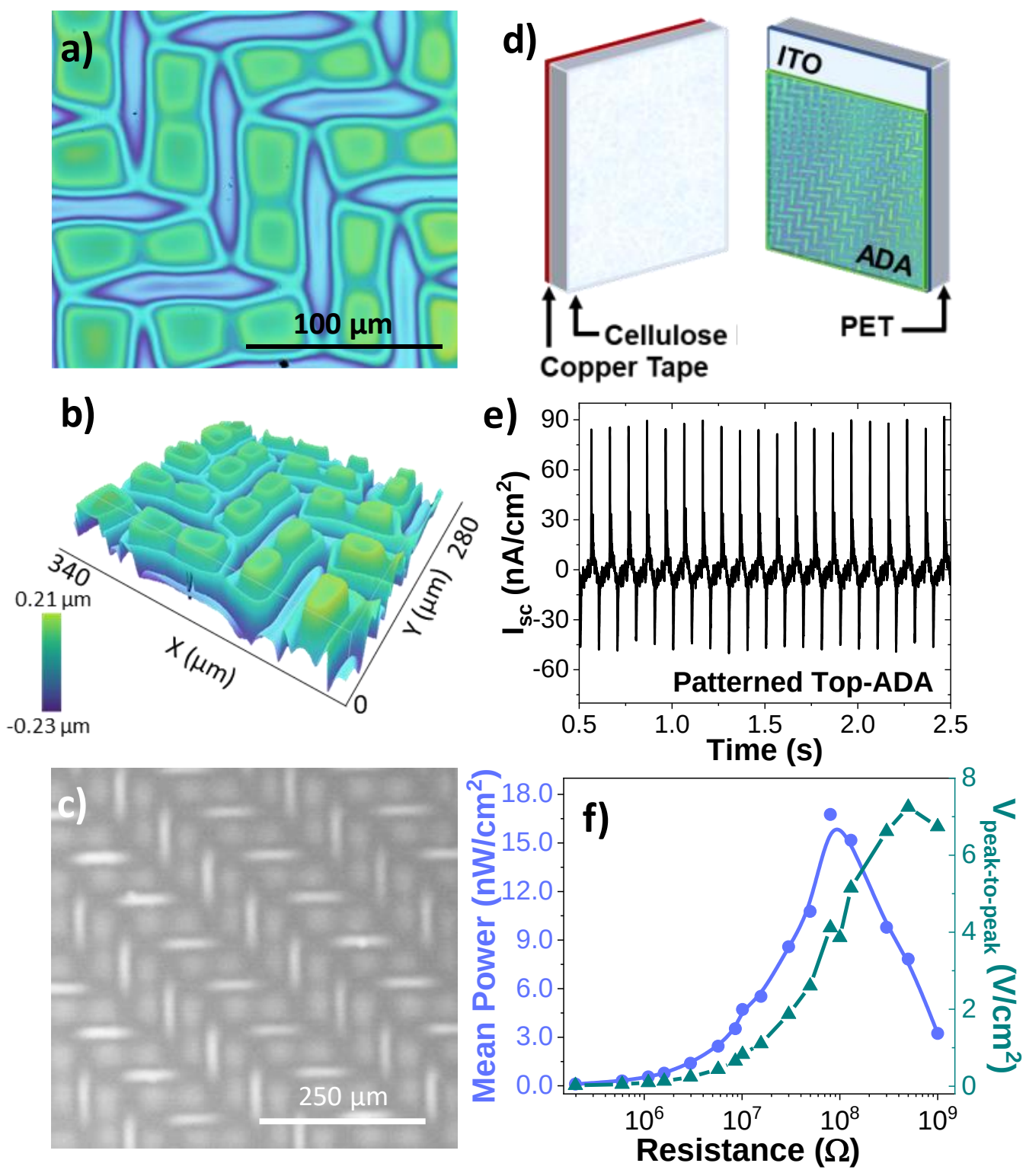

**Figure 6. Device performance with a patterned Top ADA and a cellulose counter electrode** a) Patterned Top-ADA top view confocal microscopy image and b) 3D reconstruction from confocal microscopy images c) Patterned Top-ADA top view by SEM on Si(100). d) TENG device schematic e) Short circuit current output vs time f) Mean power and peak-to-peak voltage for variable resistance load. All triboelectric outputs were obtained at 10 Hz and under a force of 2 N.

These factors contribute to more efficient charge transfer per cycle. Consequently, unlike previously reported systems,[62] in which buckling is induced cyclically, our buckled Top-ADA structure is pre-formed and remains stable throughout long tests. Hence, the device exhibits remarkable stability and durability during extended mechanical cycling as presented in **Figure 7**. After six hours of continuous operation, i.e., equivalent to over 100 000 actuation cycles, the output voltage remained at 91.8% of its initial value, decreasing only from 36.5 V to 33.5 V. Similarly, the average power density was sustained at 79.7% of its original level, moving from 0.79 µW cm$^{-2}$ to 0.63 µW cm$^{-2}$.

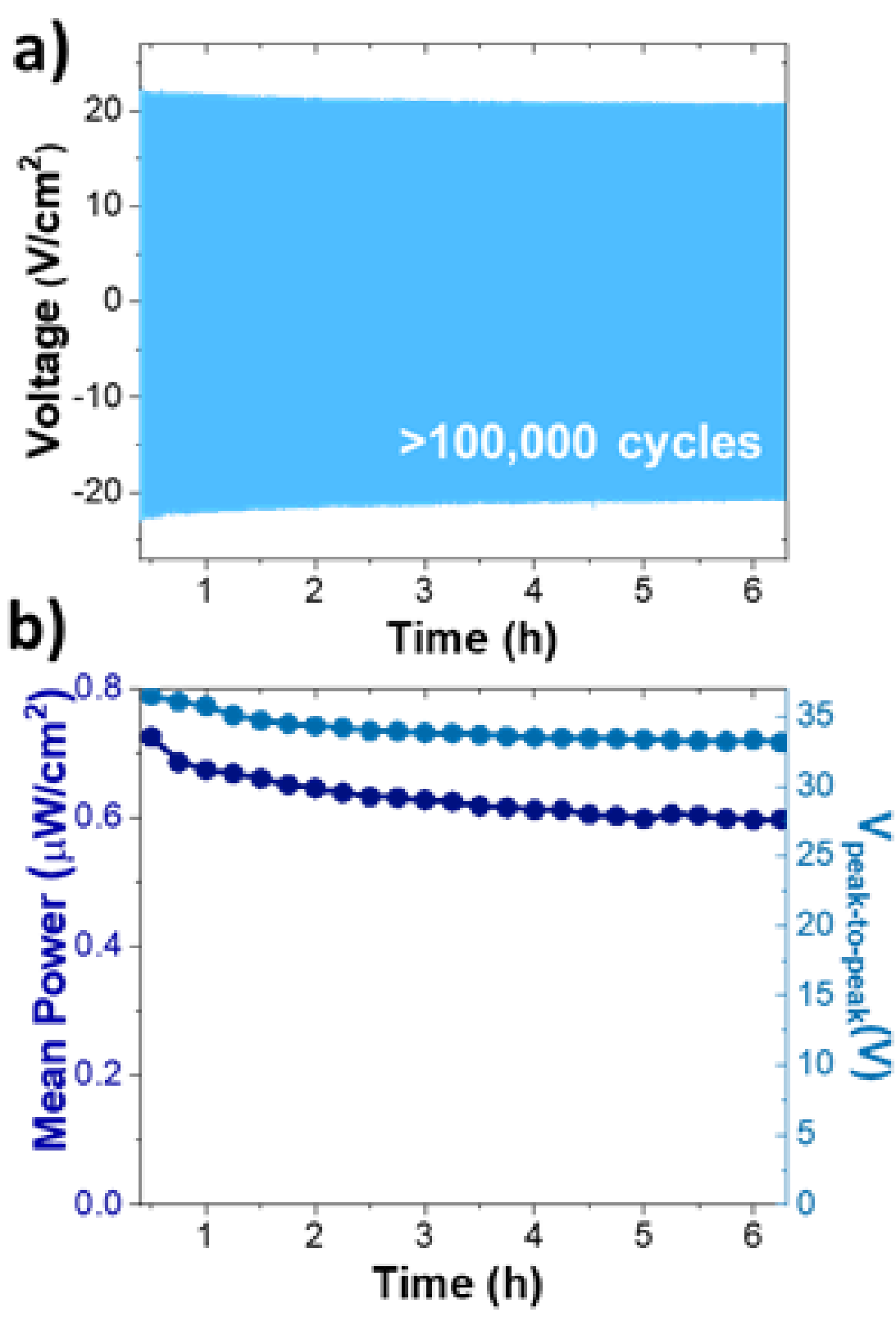

**Figure 7. Durability test for Bottom-ADA vs buckled Top-ADA TENG.** a) Voltage output at 100 MΩ. b) Maximum power density and peak-to-peak voltage. The device was subjected to sustained intermittent vertical contact at 5 Hz and 2N over 100 000 cycles.

These results confirm the high reliability and endurance of the system, validating its potential for long-term triboelectric energy harvesting without significant performance degradation. Thus, by employing the same experimental conditions, but simply varying the reactor position, it is possible to fabricate a competitive pair of triboelectric counterpart thin-film layers while avoiding fluorinated polymers.

### 2.4 Adamantane encapsulation for hybrid triboelectric-piezoelectric nanogenerators, and combination with low-dimensional nanostructures.

The hybridisation of nanogenerators and the exploitation of multifunctional nanomaterials have been proposed as reliable pathways for multi-source energy harvesting and for enhancing the performance of single-source harvesters.[64–67] Hybrid piezo-triboelectric nanogenerators (HPTNGs) are representative of this latter case. Thus, depending on the number of common electrodes, HPTNGs can be classified as mode-I (sharing both electrodes), mode-II (sharing one electrode), or mode-III (with PENG and TENG counterparts working with independent electrodes).[68] The hybrid combinations have recently been demonstrated to respond to broadband vibrations, including ultrasonic activation, and to serve as advanced, self-powered wearable sensors for monitoring vital signals or powering intelligent textiles.[67,69] A widely employed combination involves ZnO nanowires as a piezoelectric material embedded in triboelectric polymer matrices, such as PDMS or PVDF. However, the stability and properties of the ZnO surface are affected by certain solvents, such as water, alcohols, DMSO, and acetone.[70,71] In this context, exploiting plasma-processable polymers and polymer-like layers offers advantages beyond their direct compatibility with ZnO and other organic and inorganic piezoelectric fillers, namely, high conformality, fine control of composition gradients and layer thickness, and advanced interface engineering options to enhance adhesion with top and bottom electrodes.[72] In this section, we have explored the use of ADA triboelectric layers in mode-I HPTNGs, in combination with hydrothermally grown ZnO nanorods (NRs).

Notably, the total thickness of the ADA layer can be easily tuned by controlling the deposition duration. **Figure 8** compares the growth of a thick layer of Top-ADA (~3000 nm, in panel a, with a thin Bottom-ADA layer (ca. ~500 nm) on ZnO nanorods in panel e. As shown in the SEM cross-section micrographs, the Bottom-ADA layer exhibits greater conformality during deposition, yielding a shell that fully covers the rods and extends to the substrate interface. In the Top-ADA layer, the polymer continues to grow over the nanorods, ultimately forming a flat surface. In both cases, the ADA layer completely covers the ZnO NRs. It is interesting to note that on the ZnO NRs substrate, the Top-ADA grows smoothly, without the buckling observed on ITO-PET substrates. The schematics shown in Figure 8 illustrate two configurations of adamantane polymer on the nanorods, coupled with a PFA counter surface, and assembled into an intermittent vertical-contact triboelectric nanogenerator. **Figure S6** presents equivalent experiments for the aluminium countersurface. The piezo-triboelectric performance of the systems has been characterised for different frequencies. A first result is that in both cases, although much more pronounced for the Top-ADA device, the higher the frequency, the higher the mean power and voltage output (see panels b and c for Top-ADA and f and g for Bottom-ADA). This result agrees with previous reports in the literature, which show a slight enhancement in power output with increasing frequency in similar ranges.[73]

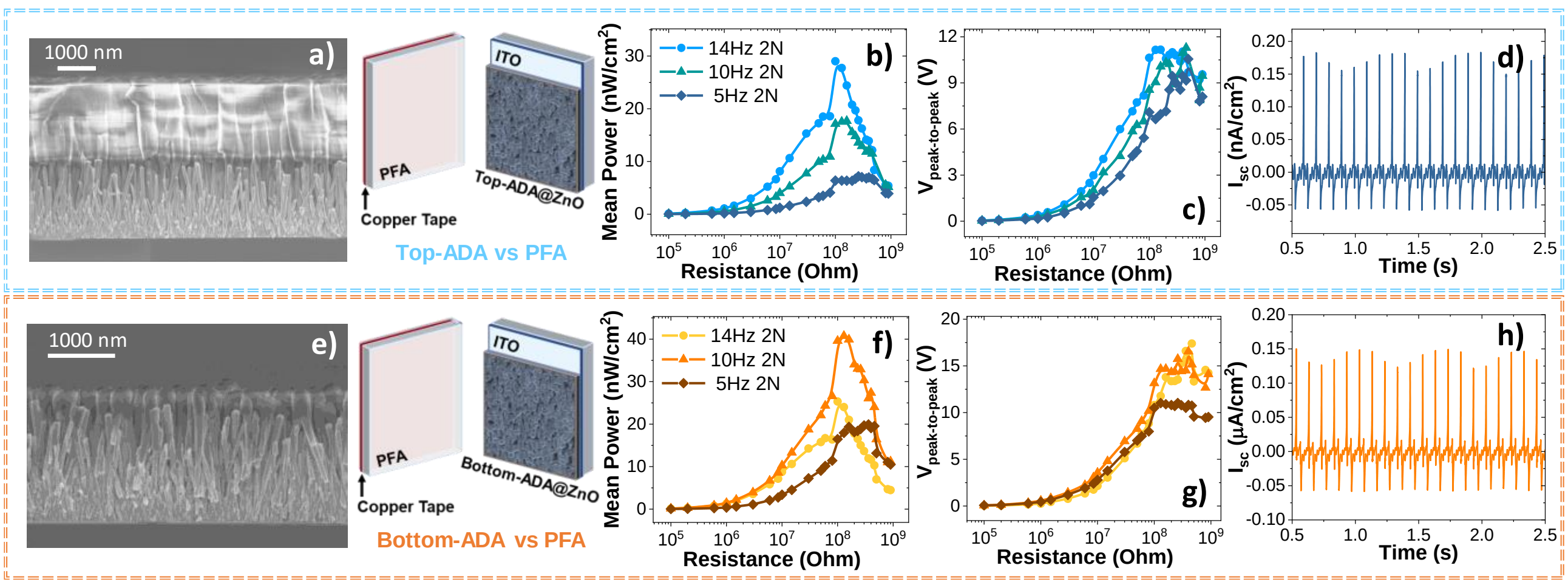

**Figure 8. Mode I hybrid piezo – triboelectric nanogenerators based on ADA plasma polymers.** a) and e) characteristic cross-section SEM micrographs for ZnO nanorods coated with Top-ADA (~3000 nm thick) (top side of the figure) and Bottom-ADA (~500 nm thick) (Bottom) layers, respectively. The schematics show the different configurations for the hybrid systems coupled with PFA. b) and f) Mean power vs resistance loads, and c) and g) peak-to-peak voltage vs resistance loads characterised for different activation frequencies and forces as labelled; d) and h) maximum short circuit current vs time curves.

A more interesting result is that the mean power of the Bottom-ADA device is much higher than that of the Top-ADA interface (note that the trend is reversed when coupled against aluminium, Figure S6). Based on the results in the previous sections, the higher tribopositive (tribonegative) character of Bottom-ADA (Top-ADA) would be more favourable for contact electrification with PFA (aluminium), a tribonegative (tribopositive) surface. In addition, the piezoelectric character of the ZnO NRs would enhance the effect by additional polarisation [74] in the same direction under vertical stress[73] when paired with PFA. The outstanding mechanical properties of these adamantane polymers are also reflected in these results, as a layer as thin as ~500 nm produces robust hybrid low-dimensional materials capable of withstanding thousands of cycles, maintaining their electrical performance, and improving the ZnO NRs adhesion to the ITO-PET substrates.

### 2.5 Drop-TENG based on adamantane triboelectric surfaces.

Adamantane layers exhibit partial hydrophobicity and high stability under moisture, dripping, and freezing conditions. Thus, we have previously exploited this feature to encapsulate perovskite solar cells against moisture and water[33,75] and to develop 3D anti-icing nanofabrics.[32]

This section explores the implementation of ADA thin films as triboelectric surfaces to harvest kinetic energy from falling drops, i.e., in drop-triboelectric nanogenerators (DTENGs). DTENG layouts have recently evolved from single-electrode to top-bottom electrode switching configurations,[44,76] with an appealing increase in power-conversion density in DTENG arrays.[77–79] Herein, we have fabricated a switch electrode device following the reference article by Zuankai Wang et al., published in 2020.[34] The schematics in **Figure 9** represent the sandwiched ADA triboelectric layers deposited between an extended FTO bottom electrode and a narrower Ti/Au top electrode (see Experimental section). According to previous results, two thicknesses have been tested for the triboelectric layers: ~1.8 µm for Top-ADA and ~500 nm for Bottom-ADA. The static water contact angles of these surfaces range from 73° to 83° across different droplet compositions, including Milli-Q, rainwater, and salty droplets (see **Table S3** and **Figure S7**). Figure 9a-c and e-g gathers the characteristic voltage peaks generated by both ADA layers for three types of droplets, for the indicated resistance load ($R_L$) and a droplet volume of 37 µl (see additional characterisation in **Figure S8**). The curves show a sharp, narrow positive peak followed by a shallower, broader negative branch, corresponding to the droplet contacting the electrode layer and the subsequent drop recoiling.[77,80] The Top-ADA devices produce higher peak intensities for both milli-Q water and rainwater, ranging between 2.5 and 3 V, but show a reduced response to salty droplets. These results are in good agreement with previous reports in the literature for a similar electrode architecture.[81,82] In contrast, the Bottom-ADA devices demonstrate stable voltage generation ~2 V for all three types of droplets (see Figure 9e-g), exhibiting a slightly higher response to salty droplets, and depicting a sharper positive peak with relatively lower contribution from the recoil. These preliminary trends have been contrasted in the power versus load curves presented in panels d and h of the figure, including statistical analysis performed with the NanoDataLyzer software,[83] with at least 10 impact events per point.

Interestingly, the optimal resistance load for both devices and drop composition is in the range of hundreds of ohms, significantly lower than the values reported in the literature for equivalent layouts.[84] It is also worth noting that the maximum peak power for salty droplets obtained with the Bottom-ADA devices reaches a value of 0.2 mW $cm^{-2}$. We hypothesise that the reduced thickness of the ADA layers compared to the bulky standard polymers employed in previous studies[77,80] and their enhanced permittivity are responsible for this improved behaviour.

In addition, we have also performed several experiments to establish the optimal drop velocities and dripping frequencies (Figure 9i-j). The peaks observed for Top-ADA follow the expected trend reported in the literature.[34,44] In particular, the maximum appears at the falling height just before the higher velocity of the droplets produces their splitting. In the second regime, the reduced contact area, or even the absence of direct impact, limits interaction between the drops and the electrode, resulting in lower voltage output. Nevertheless, what is most interesting about this characterisation is the highly stable response to salty and rainwater over long-term impact periods, as shown in Figures 9j and k. The response of Bottom-ADA surfaces to salty droplets is particularly noteworthy. To illustrate this, panel k compares the output voltage of a 10 µm thick PDMS device with that of the Bottom-ADA system. The first part of the graphs shows the impact of milliQ droplets (black lines), while the second part (orange lines) shows the effect of salty droplets. In the case of PDMS, the presence of salty droplets rapidly decreases the output voltage. In contrast, the Bottom-ADA maintains a constant voltage even after hundreds of impacts. The reduction in voltage during prolonged interaction with negatively charged droplets is attributed in the literature either to the surface saturation of the triboelectric materials or to the formation of salt crystals and precipitates.[77] The tribopositive nature of Bottom-ADA enhances its response to electrophilic droplets, and its stable wettability, combined with extremely low roughness, facilitates the easy removal of any precipitates that may form. These early findings pave the way for the development of tribopositive materials specifically designed for ocean energy-harvesting applications, where saline environments are prevalent, with the additional advantage of avoiding fluorinated compositions.[85]

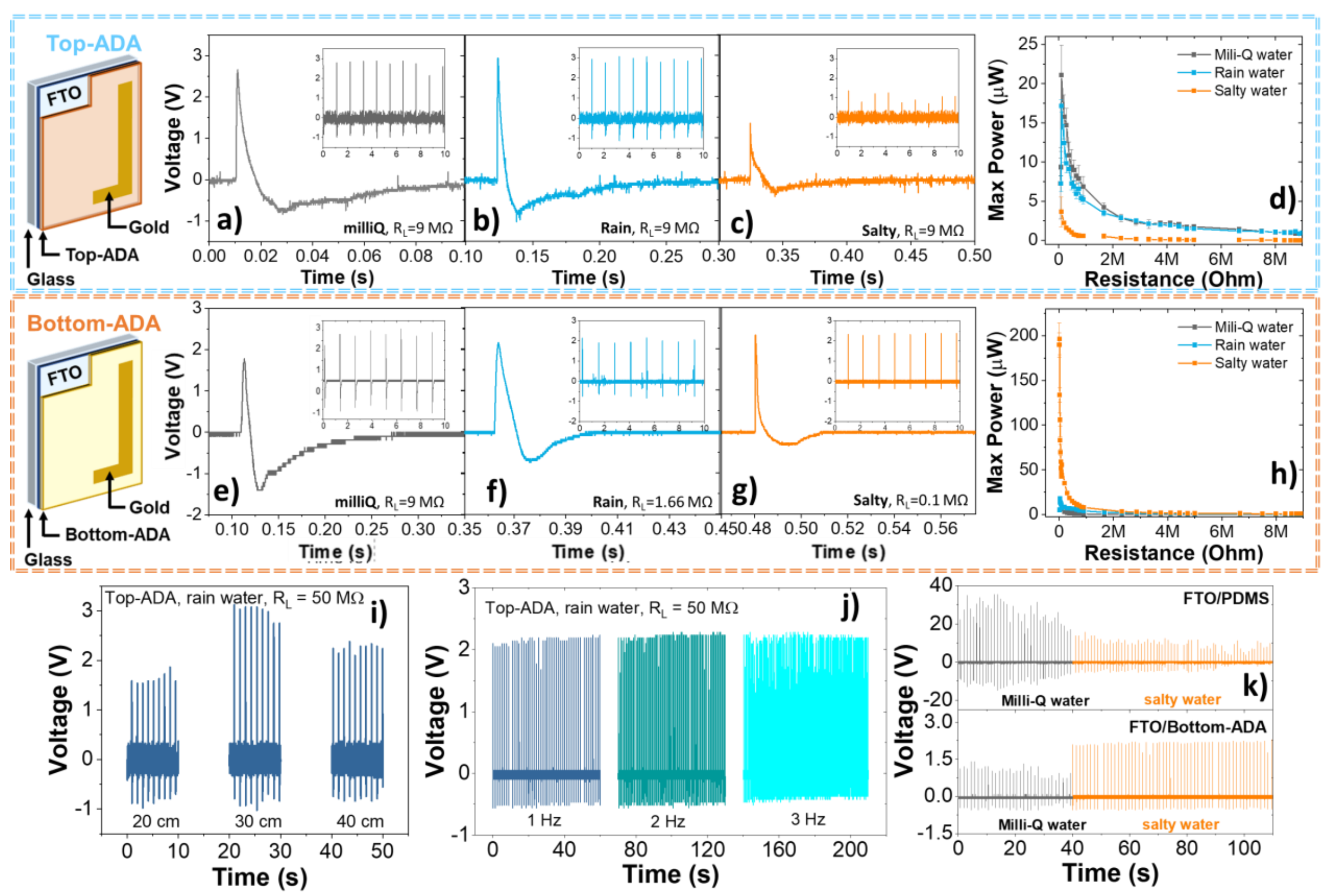

**Figure 9. Thin film Drop-TENGs enabled by plasma adamantane layers.** Characteristic V-t curves for milliQ, rain, and salty droplets of DTENGs assembled in switch electrode configuration (see schematics) corresponding to Top-ADA (1800 nm) (a-c) and Bottom-ADA (500 nm) (e-g) layers; d) and h) Maximum power for impacts of 37 µl droplets as a function of the load resistance for Top-ADA and Bottom-ADA devices, correspondingly; i) Variation of the voltage output as a function of the height of the rainy droplets for a Top-ADA device; j) Variation of the voltage output as a function of the salty droplets impact frequency for a Top-ADA device; k) Comparison of prolonged response to salty droplets for a reference PDMS device and the Bottom-ADA system.

### 2.6. Holistic Understanding of Triboelectric Mechanisms in ADA Polymers.

In this section, we aim to elucidate the working mechanism of the ADA polymer as a triboelectric material. For this purpose, we must consider the different experiments presented to date: solid-solid TENGs, hybrid systems, and liquid-solid TENGs. We will also distinguish between smooth and buckled layers.

*Smooth thin film configuration.* In the first experiments (Figure 2), Smooth Top-ADA and Bottom-ADA were tested against cellulose, showing that both exhibit tribopositive behaviour, albeit to different extents. Top-ADA produced a higher signal, indicating a weaker tribopositive character compared to Bottom-ADA, while maintaining the same polarity. This result aligns with the data presented in Figure 3 (see also Figures S2 and S3), in which both ADA-based polymers were tested against ITO (a material positioned in the middle of the triboelectric series), and Bottom-ADA showed a slightly higher output. In recent reports, Sutka et al. have demonstrated that, beyond classification into the triboelectric series, slight variations in surface roughness or mechanical properties of chemically equivalent materials are sufficient to trigger friction-driven charge generation between two surfaces.[86] Thus, these authors have demonstrated enhanced power conversion in bilayer PDMS laminates defined by elastomer cross-linking or surface roughness. Wang et al. hypothesised that electron transfer is the primary mechanism of contact electrification between two similar surfaces, driven by slight differences in surface roughness and charge transfer triggered by flexoelectricity.[63,87] Taking into account these previous considerations and the results in Figure 1 and Sections 2.1 and 2.2, we propose that the observed differences in triboelectric character arise from the combined influence of subtle variations in RMS roughness (1.5 nm vs 0.6 nm for Top- and Bottom-ADA, respectively), differences in Young's modulus, hardness and dielectric constant (both higher for the Top-ADA layers), and, most importantly, the markedly lower $E_1$ value measured for the less crosslinked Bottom-ADA films (Figure 1b), and higher contact potential difference for Bottom-ADA defined by KPFM (Figure 1g). Since contact electrification involves interfacial electron-transfer processes governed by the energetic accessibility of surface electronic states, a lower first crossover energy indicates a reduced effective barrier for electron extraction and, therefore, an enhanced electron-donor character. This interpretation is consistent with a more tribopositive behaviour of Bottom-ADA, as materials that more readily donate electrons tend to become positively charged upon contact or friction. In contrast, the significantly higher $E_1$ observed for the Top-ADA sample suggests stronger electron localisation and trapping, limiting electron donation and shifting the triboelectric response accordingly. These results indicate that, when comparing smooth films, the higher electron-donating ability of the Bottom-ADA polymers relative to Top-ADA dominates over the differences in mechanical properties in determining the triboelectric behaviour of these materials.

*Buckled vs Smooth configuration.* Figure 5 presents the output obtained from testing Bottom-ADA against both PFA and buckled Top-ADA. Remarkably, replacing PFA with Buckling Top-ADA not only preserves the mean power density of the device but also reverses the polarity of the signal (Figure 5d-e, see also Figures S2-S4 for the entire triboelectric series). This polarity reversion points to a complex interaction in which the microcavities of the buckled film interface, the increase in surface roughness, and the differences in mechanical properties between the two ADA films dominate the interaction. We hypothesise that under these conditions, the mechanical properties and the presence of gaps become more influential in determining the charge distribution. According to the study by Sutka et al.,[88] softer polymers tend to accumulate more triboelectric charge and are more likely to be negatively charged if contacted with a harder polymer. Bottom-ADA, being the softer material, acquires a negative charge, while the contact surface of the buckled Top-ADA, due to its higher stiffness, becomes positively charged, resulting in an inverse polarity (**Scheme 2**).

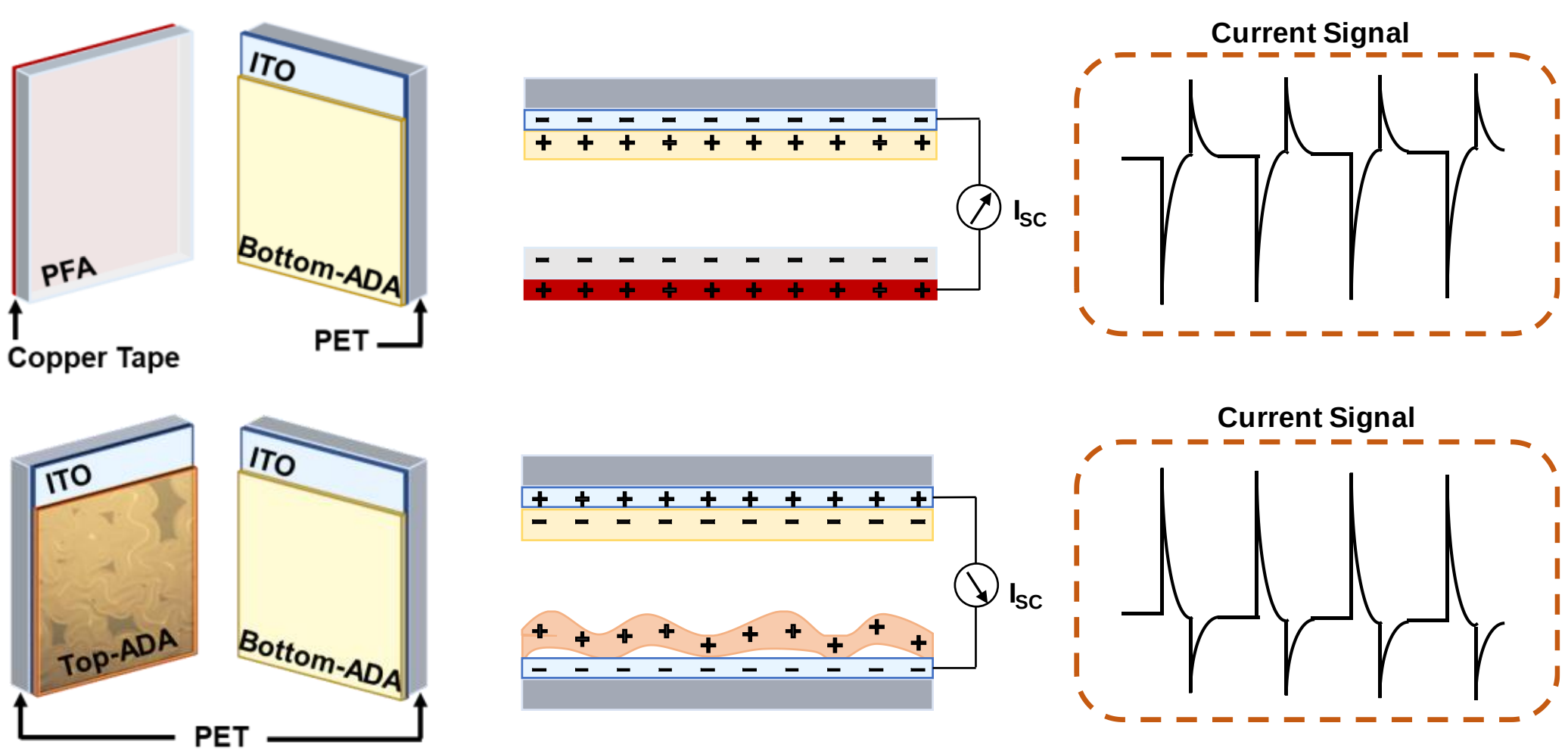

**Scheme 2.** Proposed mechanisms for triboelectrification in buckled adamantane polymer layers.

Moreover, it has been reported that bubbles enhance triboelectric performance by increasing the fraction of surfaces aligned with the device's macroscopic polarity.[89] Accordingly, the buckled films partially pinned to the substrates exhibit similarities to a bubble-like morphology. Another contributing factor is the increase in effective surface area due to buckling, which results from poorer adhesion at the interface and the release of internal stress during film formation. When this oversized layer is compressed during vertical cycling, its internal surface recontacts the ITO substrate, and the additional surface very likely forms additional wrinkles that increase surface roughness. The enhanced surface roughness promotes localised stress concentrations, thereby facilitating charge transfer, reducing the energy needed for bond break, and through electron cloud overlap.[2,90] Additionally, these wrinkles may act as cleavage points, also promoting material exchange at the interface.[40] It is the same effect observed in Figure 6, where the patterned adamantane enhanced the performance of Smooth Top-ADA. In the case of the buckled Top-ADA, both effects may contribute to the enhanced triboelectric performance observed for this configuration.

*Solid-liquid configuration.* Our final experiment tested both ADA-based polymers under liquid-solid interaction conditions. When MilliQ water was dropped onto the surface, a greater triboelectrical response was observed for Top-ADA. Given that water is known to act as an electron donor and that our previous conclusion was that Top-ADA is a weaker electron donor than Bottom-ADA, this result is consistent with expectations. We propose that Top-ADA behaves as a more efficient electron acceptor, becoming the Top-ADA surface negatively charged upon interaction with the first droplets and then producing the electron double layer in the subsequent droplets, as proposed by Wang et al.[34] Conversely, when saline water was dripped onto the surface, the trend reversed: Bottom-ADA produced a stronger, more stable signal. Saline water has been reported to be a poor electron donor but a relatively good electron acceptor.[91] This behaviour aligns well with the electron-donating nature of Bottom-ADA. We hypothesise that, in this case, the saline water acquires a negative charge upon contact with Bottom-ADA, thereby enhancing triboelectric output.

## 3. Conclusions

Triboelectric charge generation is inherently complex and multifactorial, governed by the interplay among chemical composition, mechanical response, surface morphology, and environmental conditions. Here, we have reported the development of a versatile family of dielectric polymers engineered for triboelectric energy-harvesting applications. This materials platform enables operation across multiple working scenarios, spanning solid-solid contact TENGs, hybrid piezo-triboelectric systems incorporating supported ZnO nanostructures, and liquid-solid interfaces for drop-based energy harvesting.

The dielectric films are fabricated via remote plasma-assisted vacuum deposition of adamantane, yielding optically transparent coatings across the UV-vis-NIR spectral range. By tuning the plasma exposure, the resulting layers preserve a stoichiometry close to that of the molecular precursor, while exhibiting controlled degrees of precursor fragmentation and structural retention. This tunability directly translates into adjustable mechanical and electronic properties. Despite their fundamentally different growth mechanisms, the coatings exhibit relatively high Young's modulus and hardness, reaching values close to those reported for highly hydrogenated polymer-like low-density amorphous carbon coatings.[39] Thus, the RPAVD ADA coatings, compared with both standard plasma polymers, synthetic polymers, and hydrogenated amorphous carbon, represent a compromise between a relatively high Young's modulus, a fully organic composition, and complete optical transparency starting in the UV range. Furthermore, the negligible ion bombardment inherent to the remote configuration, particularly for the bottom-ADA, enables conformal deposition on sensitive substrates and molecular structures,[35] thus broadening the range of compatible device architectures and enabling the fabrication of triboelectric devices on delicate substrates, such as supported nanostructures and cellulose. This aspect is currently under investigation in our laboratories.

The triboelectric response of the films is strongly dependent on growth conditions. Bottom-ADA films display a clear tribopositive character, whereas Top-ADA films exhibit a more intermediate position within the triboelectric series (validated by the KPFM surface potential). Furthermore, controlled formation of stable buckling structures in Top-ADA layers provides an effective strategy to enhance device output and long-term stability, offering a potential alternative to perfluorinated polymer films typically used in high-performance TENG systems.

Secondary electron emission measurements reveal highly favoured electron emission, with a first crossover energy as low as 11.1 eV. This low threshold provides insight into the surface electronic structure of the plasma polymers and may help rationalise their charge-transfer behaviour during triboelectric operation. While secondary electron emission and triboelectric charging are typically treated independently, both depend on the electronic structure and charge dynamics of the near-surface region. Differences in secondary electron emission may reflect variations in surface states and defect landscapes that are also relevant to contact charging phenomena. Nevertheless, direct experimental correlations between SEY and triboelectric behaviour for polymeric systems and plasma-deposited materials remain largely unexplored. In the studied films, the preservation of the adamantane molecular cage-like structure is therefore likely to reduce the effective electron escape barrier or impart a negative-electron-affinity-like character, promoting electron loss at low energies and an enhanced tribopositive character relative to the more cross-linked structures corresponding to the Top-ADA. Based on the results, we can hypothesise that the dominant triboelectric interaction in ADA films is electron transfer.

This development addresses several critical bottlenecks in triboelectric materials technology, including compatibility with microelectronic fabrication processes, direct plasma-based deposition onto functional devices, and straightforward industrial scalability. The approach enables the fabrication of stable, durable, and high-efficiency triboelectric devices without relying on perfluorinated films or fluorine-containing polymers. In particular, the realisation of drop-based TENGs compatible with saline water is a highly relevant advancement, opening opportunities for blue-energy harvesting and operation in realistic marine environments. Altogether, these advances significantly strengthen the technological maturity of triboelectric materials and devices, accelerating their transition from laboratory demonstrations to scalable manufacturing and real-world applications in environmental energy harvesting.

## 4. Experimental Section

### *4.1. Materials.*

Adamantane powder (≥ 99% purity from Sigma Aldrich CAS: 281-23-2), PFA films (Goodfellow, 50 µm thickness), Kapton tape (from TESA 51408), ITO-PET (60 Ω/sq from Sigma Aldrich), Copper tape (3M nominal resistivity 0.005 Ω/sq), Aluminium tape (3M Aluminium Foil Tape 1436), cellulose (bleached eucalypt pulp, kindly received from Torraspapel, Zaragoza, Spain), FTO (fluorine-doped tin oxide TEC 15, resistance 15/square, 82-84.5% transmittance, substrates of 2 x 2.5 $cm^2$ coated on glass were purchased from Pilkington), silicon wafers (Topsil, 4 inches, (100), resistivity (Ω.cm)>10 000), 500 µm thickness), pure water (Merck Millipore, Milli-Q water), Acetone (VWR, 99.9%), Ethanol (VWR, 97%).

*Relative permittivity characterization*

To determine the relative permittivity of the films, a metal-insulator-metal structure was used. Thus, Adamantane thin films synthesised using Top and Bottom configurations were deposited onto conductive commercial substrates, leaving an uncoated area. Then, circular ~80 nm thick Au electrodes were deposited by thermal evaporation through a shadow mask under high-vacuum conditions (base pressure of $10^{-6}$ mbar) at a deposition rate of 0.5-1 $Å·s^{-1}$. A ~6 nm thick Ti adhesion layer was deposited immediately before gold deposition to improve adhesion. The relative permittivity of the films was studied using a QuadTech 7600 Precision Plus LCR meter from Agilent Technologies connected to a custom-built probe station. The system used a four-terminal configuration connected to the probe station by BNC cables. Measurements were programmed using a LabVIEW application from 1 kHz to 1MHz. The complex relative permittivity ($\varepsilon_r = \varepsilon_r{}' + i\varepsilon_r{}''$) was calculated according to the equations:

$$\varepsilon_r{}' = \frac{Th}{A * \varepsilon_0} * \frac{1}{\omega} (\frac{Z''}{Z'^2 + Z''^2})$$

$$\varepsilon_r{}'' = \frac{Th}{A * \varepsilon_0} * \frac{1}{\omega} (\frac{Z'}{Z'^2 + Z''^2})$$

The SEY measurements were performed on adamantane layers grown on aluminium substrates using a pulsed electron-beam method over a primary energy range from 0 to 1 000 eV. A single pulse was applied for each primary energy, with normal incidence on the sample surface. Each pulse delivered a charge of approximately 0.72 fC. The emitted secondary electrons were collected by a positively biased detector surrounding the sample, and the SEY at each primary energy was calculated as the ratio of the emitted secondary-electron current to the incident electron current. All measurements were carried out under ultra-high vacuum conditions at a pressure of $10^{-10}$ mbar to minimise scattering by residual gas molecules.

ERDA and p-EBS were performed at the Centro Nacional de Aceleradores.[92] For the p-EBS, a 2.0 MeV proton beam was used in combination with a PIPS detector set at 165° to collect backscattered ions. For ERDA, a helium beam of 3.0 MeV was employed to produce the recoil of H present in the sample. Particles were measured with a PIPS detector set at 44° with a 13 µm Mylar filter in front of the detector to avoid the arrival of helium ions. All spectra were then simulated using the SIMNRA code.[93]

Raman spectroscopic characterisation was carried out on a Horiba Jobin-Yvon LabRAM spectrometer equipped with a confocal microscope and a 50× objective, using a 532 nm green laser. The spectral resolution for this configuration was ~1.7 $cm^{-1}$. No polarisation was applied during the experiments. Low laser powers were utilised to prevent local heating.

The surface potential distribution of the ADA thin films was characterised via Kelvin Probe Force Microscopy (KPFM) using a Park NX-10 Atomic Force Microscope (AFM) system. Measurements were conducted in sideband KPFM mode to ensure high spatial resolution and sensitivity. KPFM was performed by using a conductive Pt/Ir coated tip with a spring constant of 3 N/m. The relative humidity during the measurements was ~80 %. The potential maps were processed using Park XEI software to extract the contact potential difference values.

Young's modulus and hardness values were determined using a NanoXP nanoindenter (MTS Nanosystems) equipped with continuous stiffness measurement (CSM) and dynamic contact module (DCM) capabilities. All experiments were performed at room temperature employing a diamond Berkovich indenter. The load–displacement curves were analysed following the Oliver–Pharr method[94] to extract hardness and elastic modulus as a function of penetration depth. The measurements were performed on ~500 nm thick Top-ADA and Bottom-ADA adamantane plasma polymer films deposited on Si(100) substrates. To minimise substrate influence, the mechanical properties were evaluated within a depth range limited to 10–15% of the total film thickness.

*Water Contact Angle Characterisation*

A water contact angle (WCA) measurement system, OCA20 from DataPhysics, was used to measure and compare the hydrophobicity of the ADA films (see Table S3 and Figure S7). The WCA system consists of a peristaltic pump (Perimax 12/4-SM from Spetec) to control the water flow and a set of calibrated micropipette tips that ensure a certain size of the falling droplets. 2 µL droplet of three different water conductivities (Milli-Q, rainwater and salty water) were dispensed onto the surface of the adamantane thin films considering a statistical analysis of 5 repetitions for averaged values with a margin of error below 3% of the hydrophilic surface.

*Fabrication of the different devices*

Adamantane plasma polymerized thin films were deposited by Remote Plasma-Assisted Vacuum Deposition (RPAVD). Full details of the experimental setup can be found elsewhere.[23,36] In summary, the precursor molecules were sublimated in the downstream region of an electron cyclotron resonance microwave Ar plasma (210 W, 2.45 GHz) from a heatable dispenser outside the chamber. The substrates were at room temperature (RT) and placed facing both the plasma region (Top ADA) and the precursor entrance (Bottom ADA). The sample holder was placed 12 cm below the discharge to reduce the interaction with the plasma. The growth rate was monitored in situ with a quartz crystal microbalance (QCM) placed next to the sample holder. Argon was dosed using a calibrated mass flow controller to achieve a deposition pressure of $2{\cdot}10^{-2}$ mbar. The base pressure of the reactor was $<10^{-6}$ mbar.

For the solid-solid experiments, ITO-PET was the substrate and a Kapton mask was used to prevent the polymer from depositing in the whole area and to enable easy access to the ITO, where the contacts will be made. Before the deposition, those substrates were cleaned and washed on both sides with ethanol and acetone.

Different sets of substrates and triboelectric layers were used to fabricate the counter-electrodes. For the PFA counter-electrode, a PFA film was cut into a 20 mm × 40 mm rectangle and then washed and cleaned with Milli-Q water. The PFA sheet was taped on the adhesive part of the copper tape. For the Kapton counter-electrode, a rectangle of the same dimensions was taped on the conductive part of the ITO-PET substrate. Aluminium tape was attached to the conductive side of the ITO-PET to act as a third counter electrode. Finally, a piece of cellulose film, produced by pressure filtration of a 0.2 wt% cellulose suspension, was cut to the same dimensions and pasted onto the adhesive side of the copper tape. Contacts were made on the conductive part of each electrode.

For hybrid triboelectric-piezoelectric nanogenerators device fabrication, ADA films were deposited directly on hydrothermally grown ZnO nanorods. To deposit the nanorods, a ZnO seed layer was first prepared by spin coating from a precursor solution containing 0.1 M zinc acetate ($Zn(CH_3COO)_2{\cdot}2H_2O$, Sigma-Aldrich) and 0.1 M 2-aminoethanol ($NH_2CH_2CH_2OH$, 99%, Sigma-Aldrich) dissolved in 2-methoxyethanol ($CH_3OCH_2CH_2OH$, 99%, Sigma-Aldrich). The solution was spin-coated onto 1 cm × 2 cm masked ITO-PET substrates at 500 RPM for 10 s, then at 3000 RPM for 30 s, followed by heating in air at 120°C for 15 min. The procedure was repeated 4 times. In the final step, the growing seed layer was heated to 120 °C for 30 minutes. The seeded substrates were suspended face-down in a jar containing an aqueous solution of 25 mM zinc nitrate (98%, Sigma-Aldrich) and 25 mM hexamethylenetetramine (99.5%, Sigma-Aldrich), and placed in a preheated oven at 90 °C for 5 h. The solution was then replaced with a fresh one, and the process was repeated to promote further growth of the ZnO nanorods. This procedure was carried out for a total growth time of 20 h.

For the liquid-solid experiments, ~500 nm and ~1.8 µm thick Bottom-ADA and Top-ADA films were deposited on FTO-Glass substrates, respectively. Before deposition, the FTO substrates were brushed with a Hellmanex solution in water (2:98 vol%) and then rinsed with deionised water. The substrates were sequentially sonicated for 15 min in Hellmanex solution, deionised water, isopropanol, and acetone. Finally, they were treated by UV/ozone for 15 min using an Ossila UV Ozone Cleaner. A mask was used to prevent deposition on part of the substrate, enabling easy access to the FTO (bottom electrode). The top electrode was produced by metal sputtering through a shadow mask. First, an aluminium foil was patterned with the geometry of the top electrode by laser ablation (NdYAG, 1064 nm Powerline E, Rofin-Baasel In). Then, 80nm thick Au L-shape electrodes were deposited over the triboelectric layers by thermal evaporation using a 2-3 nm thick Ti adhesion layer.[34] Finally, copper wires were connected to the Bottom and top electrodes using conductive silver epoxy, and the assembly was reinforced with a commercial sealant.

*Set up for measurements and characterisations.*

For all solid-solid experiments, a magnetic shaker (Smart Shaker K2007E01 from The Modal Shop) attached to a force sensor (IEPE model 1053V2 from Dytran Instrument, Inc.) was used to apply the tapping motion. This magnetic shaker can vary the frequency and tapping force (fixed at 2N for the experiments). The force of this mechanical stimulus was measured with a digital oscilloscope (Tektronix TDS1052B). The response of the devices was measured using a Keithley 2635A. The different loads used in the experiments were provided by a handmade resistance box. The NanoDataLyzer software was used for statistical analysis and for automated acquisition and synchronisation of the stimulus and output voltage.[77,83]

For all liquid–solid experiments, water droplets were released from a height of 30–50 cm onto the triboelectric surfaces using a 5 mL polyethylene pipette. The pipette dispensed 37 µL droplets of Milli-Q water, rainwater, and saline water through a 0.25 mm stainless-steel needle. The pipette was connected to a peristaltic pump to control the dispensing frequency and was rigidly fixed to a laboratory stand with clamps, allowing precise height adjustment. The D-TENG was fixed to a 45° tilted flatbed. The output voltage of the triboelectric surface was measured using a PicoScope oscilloscope. A high impedance probe with 10 MΩ of internal resistance was used to measure the voltage across different loads. Load impedances were varied from 50 kΩ to 99 MΩ by using a resistance box connected in parallel with the 10 MΩ resistance probe. Measurements were carried out at 20 °C and 65% relative humidity. Energy and peak power conversion vs load curves were characterised from high to low load resistance values. Data treatment was performed by statistical analysis of at least 10 drops/events per condition using the NanoDatalyzer software.

Rainwater was collected on 8 February 2024 in Seville, with a measured conductivity of approximately 6 mS $cm^{-1}$. Saline water was prepared by dissolving NaCl in Milli-Q water at a concentration of 0.585 g per 100 mL, yielding a conductivity of 60 mS $cm^{-1}$. Conductivity measurements were performed using a Mettler Toledo InLab 731-ISM conductivity probe connected to a Mettler Toledo SevenExcellence multiparameter meter.

*Calculation of average and instantaneous power densities and energy per droplet.*

Assuming that $V_{Load}$ is the measured voltage across a resistive load of $R_L$, the power is calculated as, $P_L(t) = V_L^2(t)/R_L$. The energy obtained in a cycle (called Energy per cycle, $E_{cycle}$) is calculated by integrating the power as, $E_{cycle} = \int_0^{t_{cycle}} P_L(t) \cdot dt = \int_0^{t_{cycle}} \int V_L^2(t)/R_L \cdot dt$; where $t_{cycle}$ means the time a cycle lasts. So, the Mean Power is obtained as, $P_{mean} = \frac{E_{cycle}}{t_{cycle}}$.

The software (NanoDataLyzer), developed by the authors, was used to perform statistical analysis of the mean power, energy per cycle, and peak-to-peak voltage. The software is openly available on Zenodo.[83]

**Acknowledgements**

The authors thank the projects PID2022-143120OB-I00, PCI2024-153451 and TED2021-130916B-I00 funded by MCIN/AEI/10.13039/501100011033 and by "ERDF (FEDER)" A way of making Europe, Fondos NextgenerationEU and Plan de Recuperación, Transformación y Resiliencia. Project DGP_PIDI_2024_02239 funded by Consejería de Universidad, Investigacion e Innovación of Junta de Andalucía. Project ANGSTROM was selected in the Joint Transnational Call 2023 of M-ERA.NET 3, an EU-funded network of about 49 funding organizations (Horizon 2020 grant agreement No 958174). The authors acknowledge Microscopy Services (KPFM) from CITIUS (University of Seville) and from ICMS (SEM), Wettability Characterization Service from ICMS, and Centro Nacional de Aceleradores for providing access to their facilities under the proposal CNA-STD020/26. FNG acknowledges the "VII Plan Propio de Investigación y Transferencia" of the Universidad de Sevilla. The project leading to this article has received funding from the EU H2020 program under grant agreement 851929 (ERC Starting Grant 3DScavengers). The authors further acknowledge the SUsPlast platform (CSIC). GPM thanks the MICIN/AEI for the concession of an FPI Grant (State Programme for the Promotion of Talent and its Employability in R+D+I, PRE2020-093949).

**Data Availability Statement**

Datasets will be available from the authors upon reasonable request.

# Supporting Information

**Adamantane plasma polymers: fluorine-free vacuum-processable triboelectric thin films for all-triboelectric nanogenerator configurations**

*Gloria P. Moreno-Martínez,[a] Fernando Núñez-Gálvez,[a,b] Hari Krishna Mishra,[a] Triana Czermak,[a] Xabier García-Casas,[a] Vanda Cristina Godinho,[a] Bernd Wicklein,[c] Juan Carlos Sánchez-López,[d] Javier Ferrer,[e,f] Isabel Montero,[c] Juan Ramón Sánchez-Valencia,[a] Andris Sutka,[g] Francisco Aparicio,[a] Angel Barranco,[a]* Ana Borrás[a]**

a) Nanotechnology on Surfaces and Plasma Laboratory, Consejo Superior de Investigaciones Científicas (CSIC), Materials Science Institute of Seville (CSIC-US). c/ Américo Vespucio 49, 41092, Seville (Spain).
b) Departamento de Física Aplicada I, Universidad de Sevilla, C/ Virgen de Africa 7, 41011, Seville (Spain)
c) Materials Science Institute of Madrid (ICMM), Consejo Superior de Investigaciones Científicas (CSIC), 28049, Madrid (Spain).
d) Tribología y Protección de Superficies, Consejo Superior de Investigaciones Científicas (CSIC), Materials Science Institute of Seville (CSIC-US). c/ Américo Vespucio 49, 41092, Seville (Spain).
e) Departamento de Física Atómica, Molecular y Nuclear, Universidad de Sevilla, Aptdo 1065, Sevilla, 41012, Spain
f) Institute of Physics and Materials Science, Faculty of Natural Sciences and Technology, Riga Technical University, LV-1048, Riga, Latvia

*angelbar@icmse.csic.es; anaisabel.borras@icmse.csic.es*

**Table S1.** Thin film composition determined by p-EBS and ERDA analysis of samples deposited on Si(100).

| Sample | Areal density ($10^{15}$ at/cm$^2$) | H (at %) | C (at %) | O (at %) |
|---|---|---|---|---|
| Top-ADA | 6070 ± 225 | 59.3 ± 1.2 | 35.3 ± 1.1 | 5.4 ± 1.3 |
| Bottom-ADA | 5740 ± 206 | 61.0 ± 1.2 | 36.9 ± 1.2 | 2.1 ± 0.5 |
| ADA molecular formula | -- | 61.5 | 38.5 | 0 |

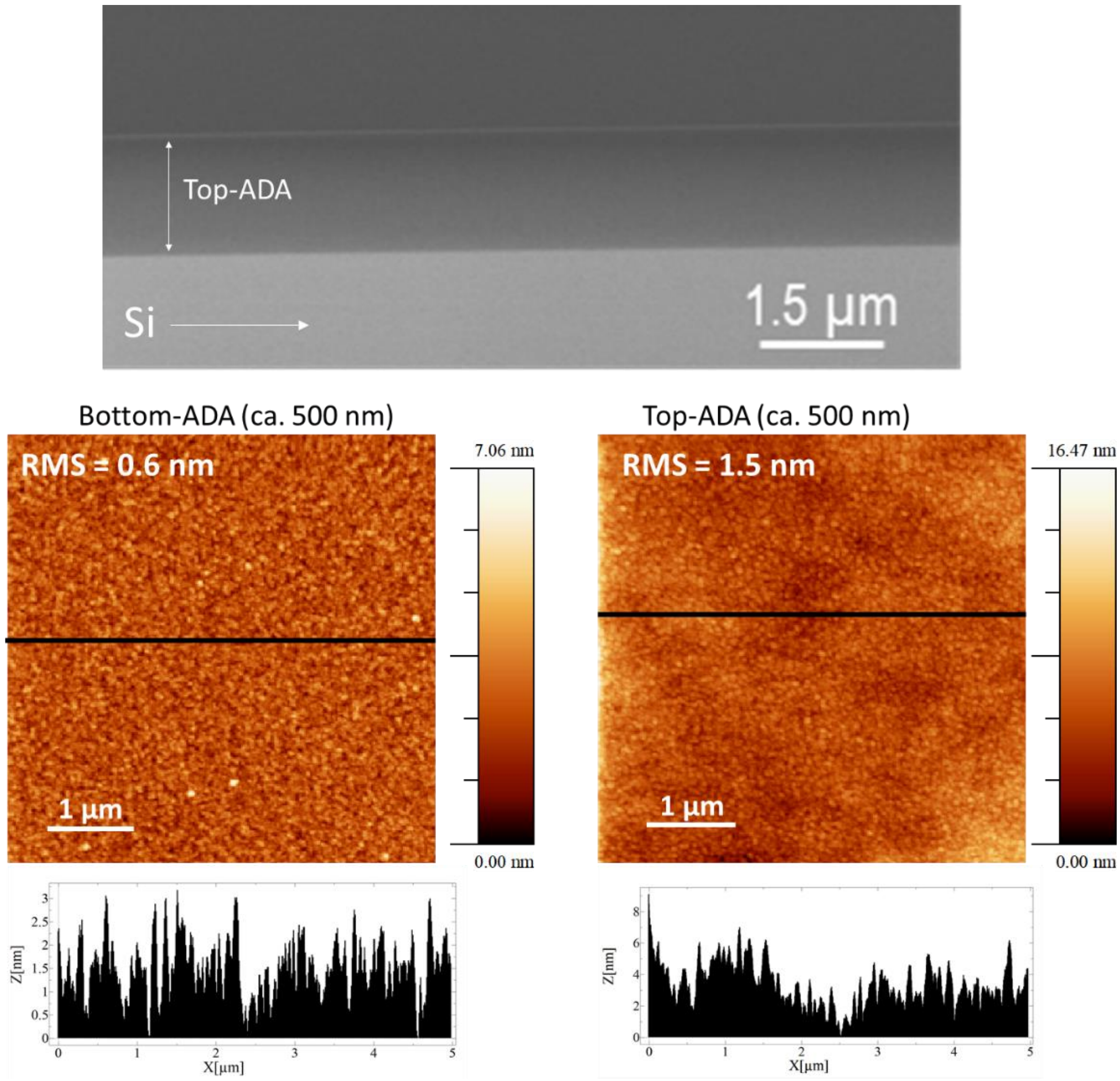

**Figure S1.** Top) cross-sectional SEM micrographs for Top-ADA film deposited on Si (100). Bottom) AFM characterisation of Bottom-ADA (left) and Top-ADA (right). Both layers were fabricated with a thickness of ~500 nm on silicon wafers, and the micrographs were taken in tapping mode. The figure includes the RMS value and characteristic line profile for both surfaces.

**Table S2.** Young's Modulus and hardness characterisation for samples of 500 nm thick deposited on Si(100) determined by nanoindentation.

| | **Young Modulus** (Average) E (GPa) | **Standard Deviation** | **HARDNESS** (Average) H (GPa) | **Standard deviation** | **Indentation Depth (nm)** | **Valid Indentations** | **Evaluation Range (nm)** |
|---|---|---|---|---|---|---|---|
| **Bottom-ADA** | 7.64 | 0.23 | 0.57 | 0.03 | 50 | 12 | 15-30 |
| **Top-ADA** | 10.24 | 0.27 | 0.84 | 0.04 | 50 | 25 | 20-25 |

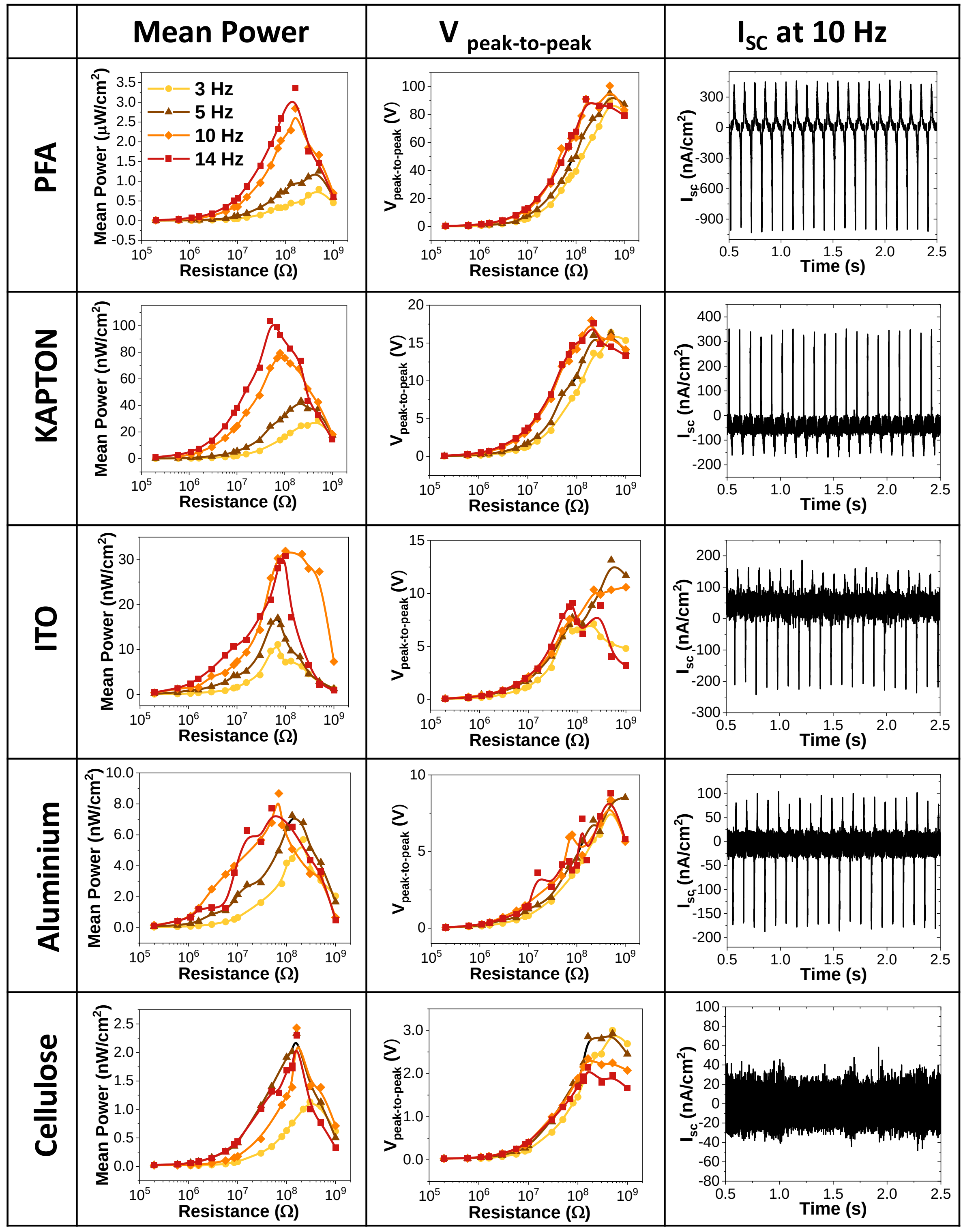

**Figure S2.** Triboelectric performance for Bottom-ADA (~450 nm) against different triboelectric surfaces for varying frequencies and at 2 N.

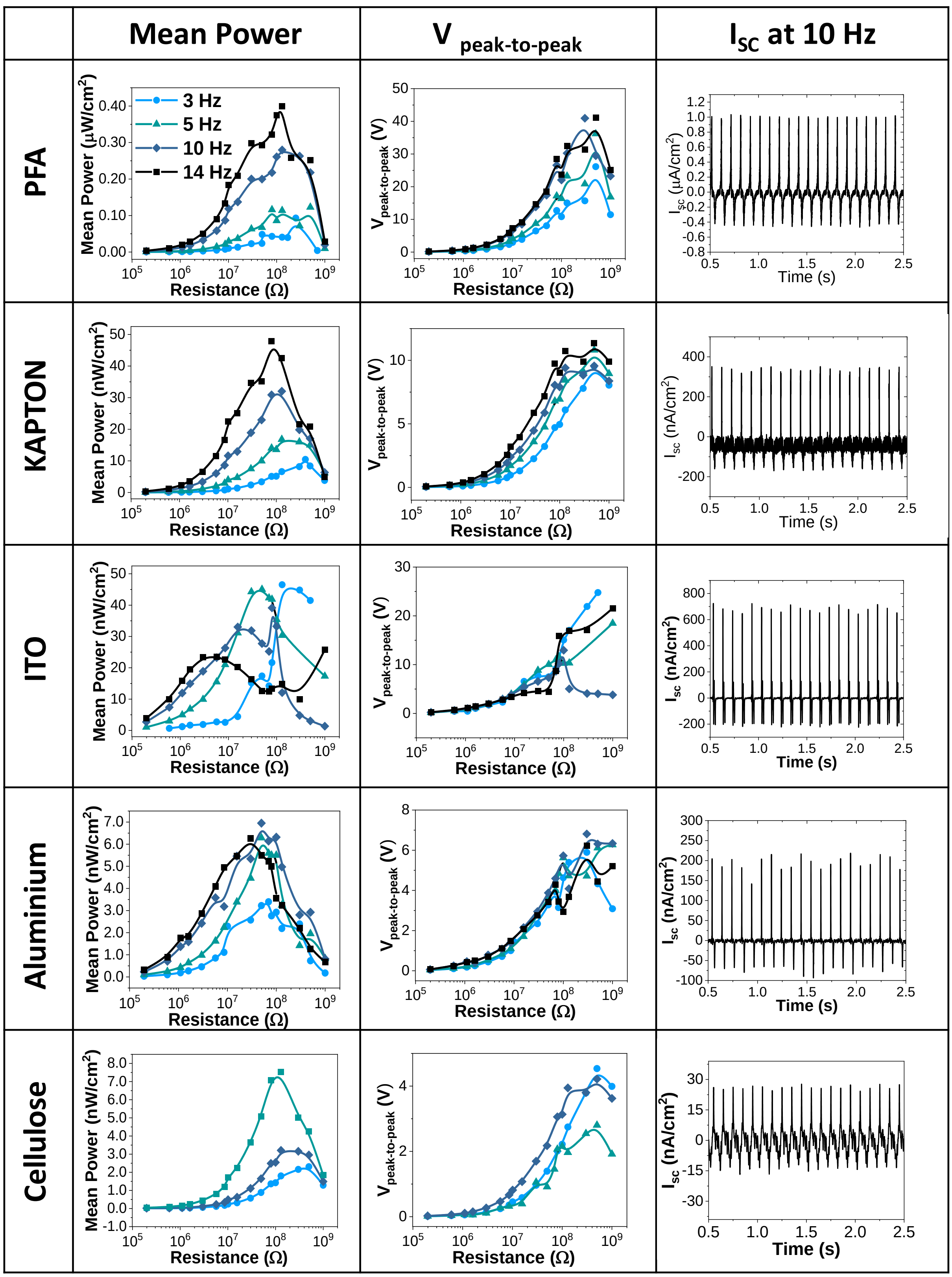

**Figure S3.** Triboelectric performance for Top-ADA (~500 nm) against different triboelectric surfaces for different frequencies and at 2 N.

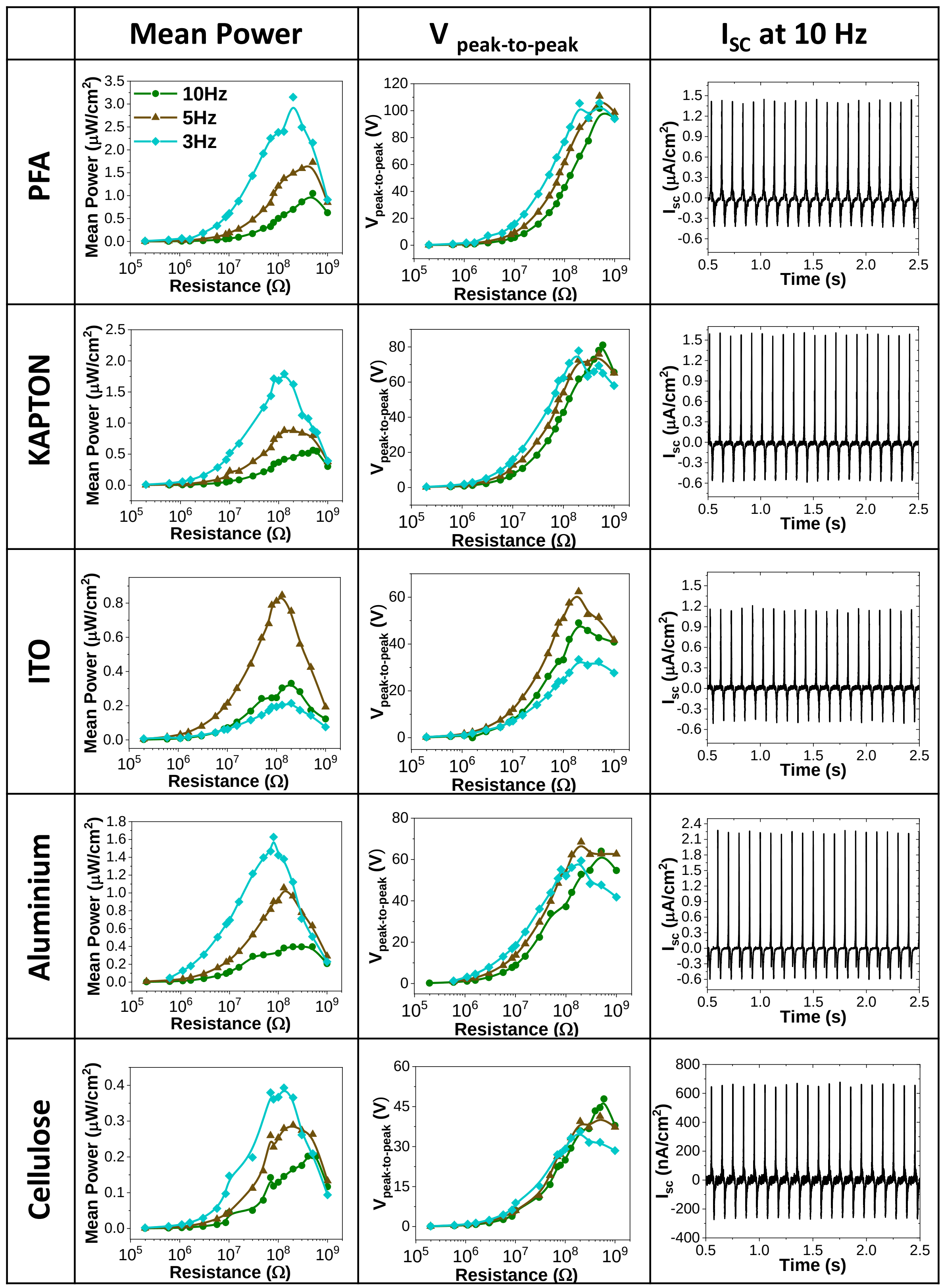

**Figure S4.** Triboelectric performance for buckled Top-ADA (~2.8 µm) against different triboelectric surfaces for varying frequencies and at 2 N.

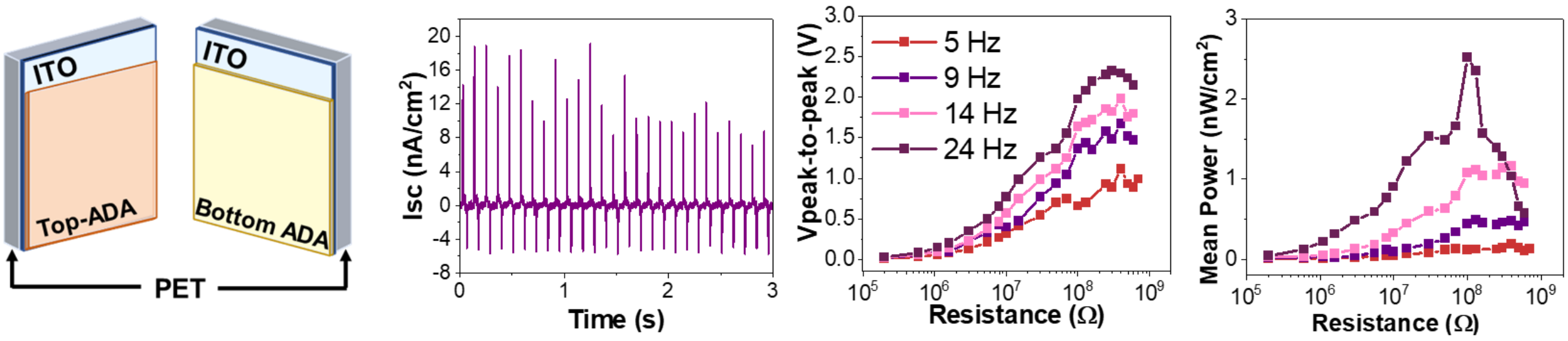

**Figure S5.** Triboelectric output for smooth Top-ADA (~500 nm) vs Bottom-ADA (~450 nm) for varying frequencies and at 3 N.

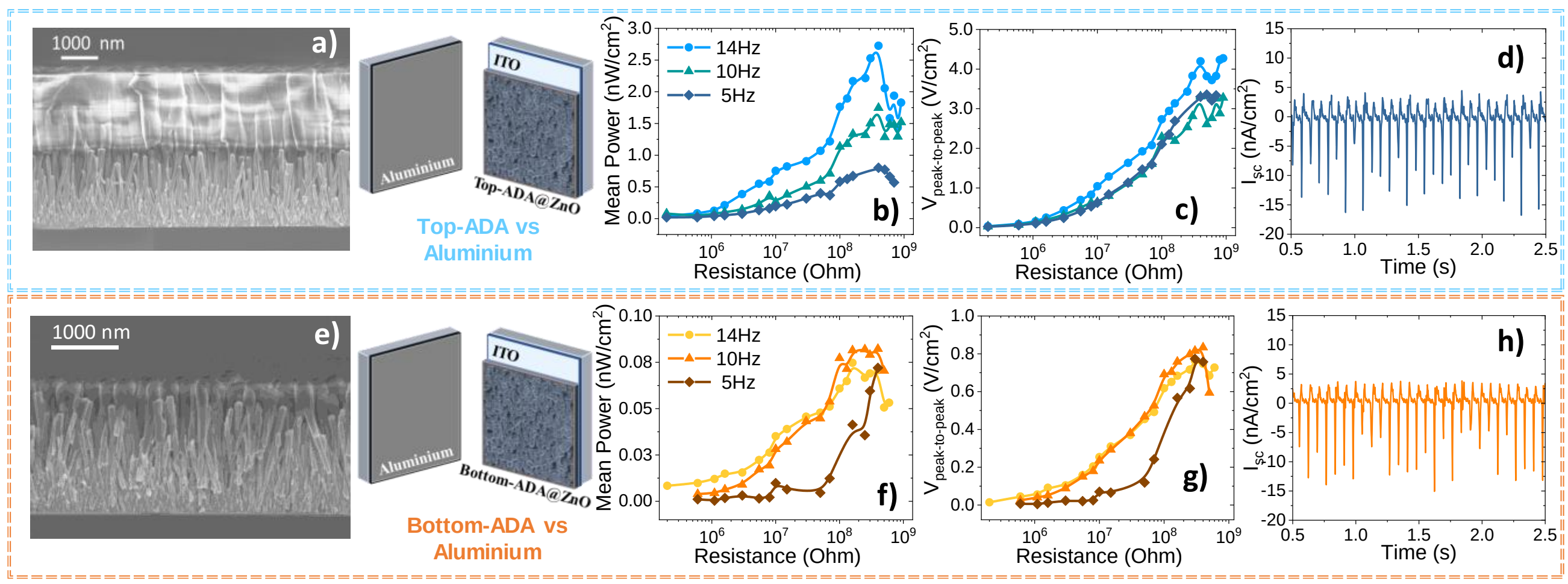

**Figure S6. Mode I hybrid piezo-triboelectric nanogenerators based on ADA plasma polymers.** a) and e) Characteristic cross-section SEM micrographs for ZnO nanorods coated with~ 3µm thick Top-ADA film and ~500 nm thick Bottom-ADA film, respectively. The schematics show the different configurations of the hybrid systems using aluminium counter electrodes. b) and f) Mean power vs resistance loads. c) and g) Peak-to-peak voltage vs resistance loads characterised for different activation frequencies and forces as labelled. d) and h) Maximum short circuit current vs time curves.

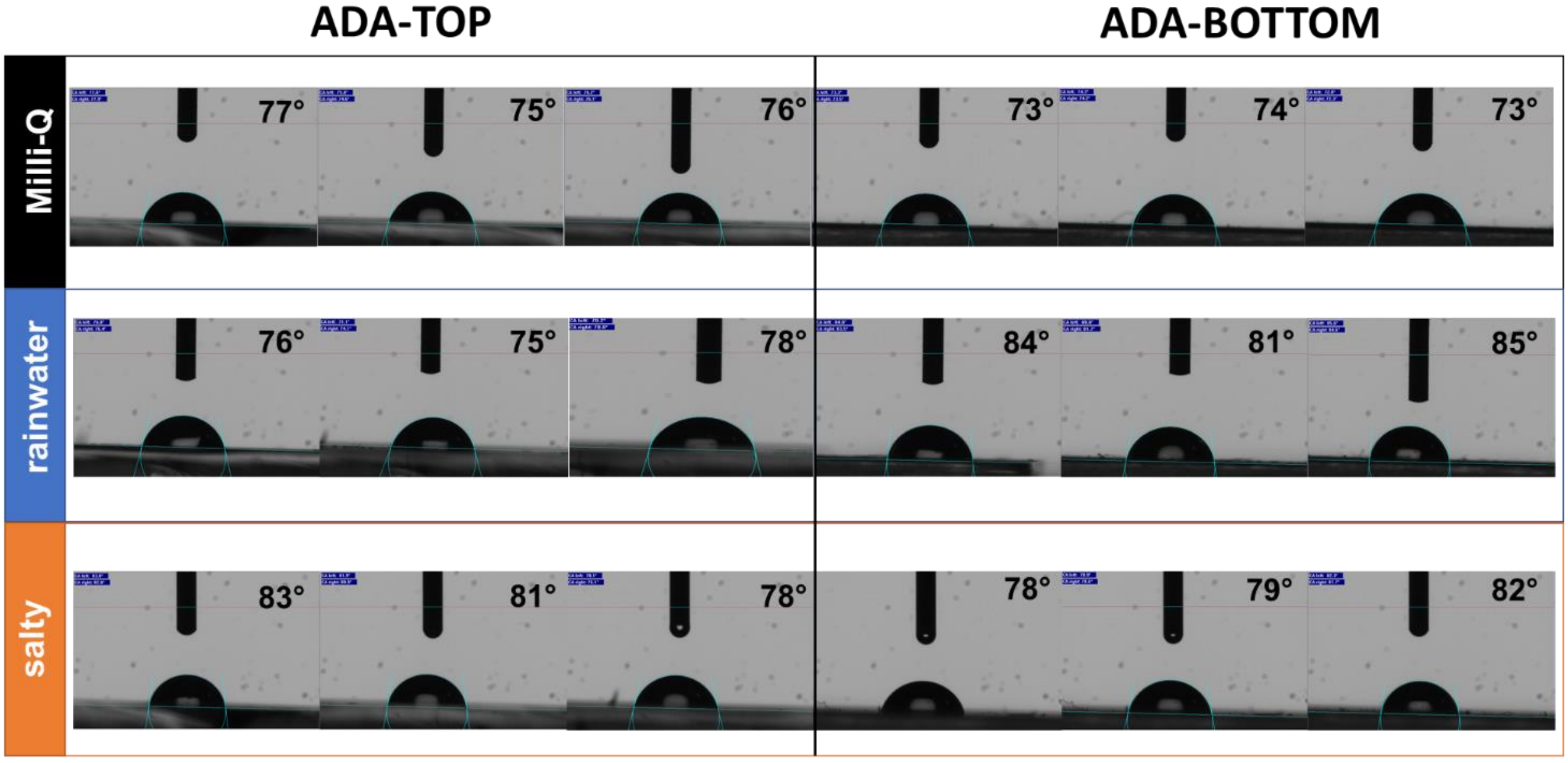

**Figure S7.** Characteristic water contact angle for the two ADA surfaces for milli-Q, rainwater and salty water, for a droplet volume of 2 µL.

**Table S3.** Static water contact angles for ADA-Top and ADA-Bottom layers to three different droplet compositions (drop volume was settled at 2 µL).

| | **Bottom-ADA** | | | **Top-ADA** | | |
|---|---|---|---|---|---|---|
| | milliQ | Rainwater | Salty water | milliQ | Rainwater | Salty water |
| WCA (º) | 73 ± 1 | 83 ± 2 | 80 ± 2 | 76 ± 1 | 77 ± 1 | 82 ± 1 |

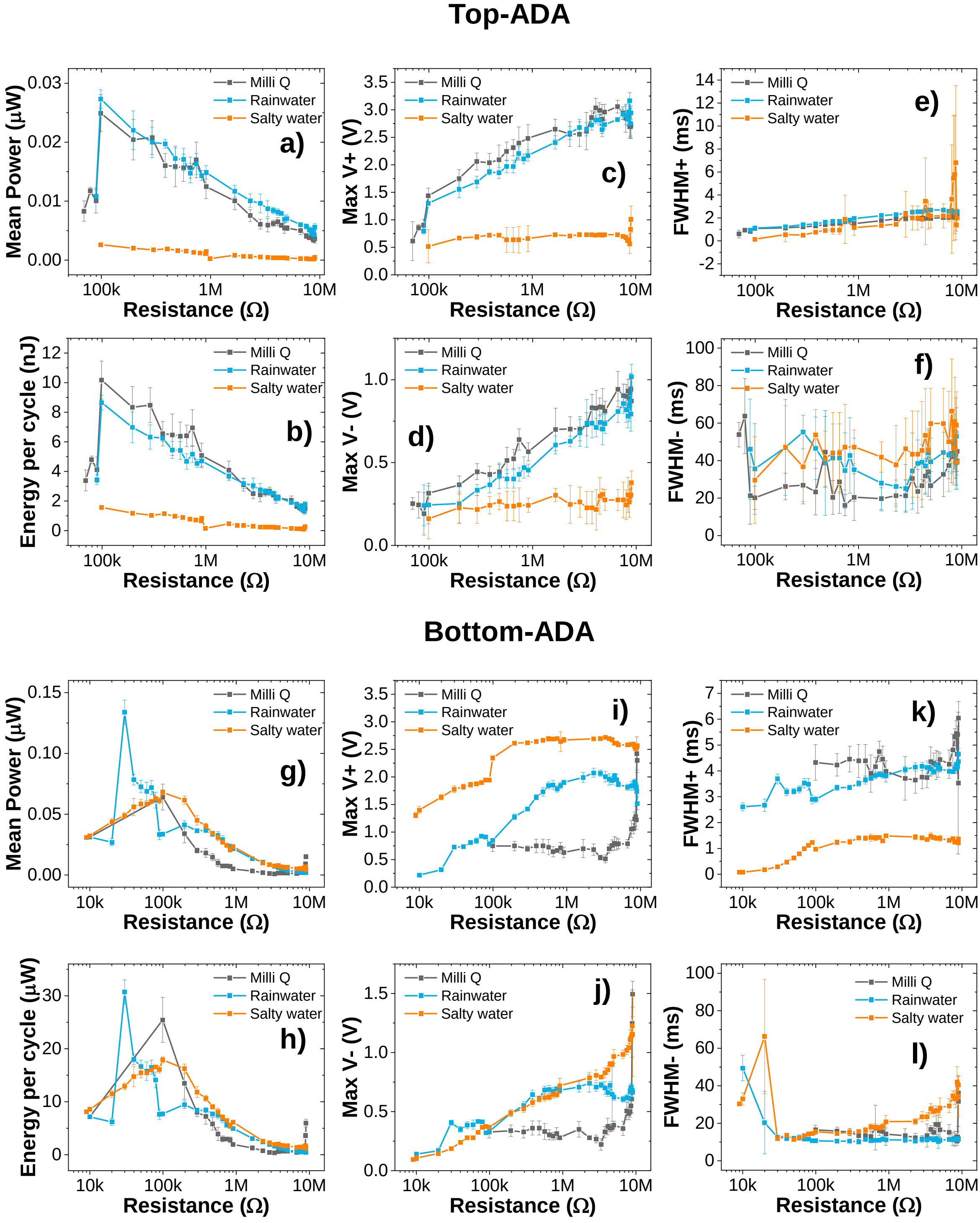

**Figure S8**. Drop-TENG output parameters as a function of $R_{Load}$ for the water-droplet characterisation of Top-ADA (a-f) and Bottom-ADA (g-l). a) and g) Mean power (full signal integrated area). b) and h) Energy per cycle (considering a cycle as an event). (c-d) and i-j) Maximum V+/V- output peak power. e-f) and k-l) Full Width at Half Maximum (FWHM+/FWHM-).